\documentclass[a4paper,onecolumn,unpublished,floatfix,11pt]{quantumarticle}
\pdfoutput=1
\usepackage[utf8]{inputenc}
\usepackage[english]{babel}
\usepackage[T1]{fontenc}
\usepackage{amsmath}
\usepackage{hyperref}

\usepackage{tabularx}
\usepackage{makecell}

\usepackage{tikz}
\usepackage{lipsum}

\usepackage{mathtools}
\usepackage{physics}
\usepackage{textgreek}

\usepackage{soul}

\usepackage{svg}
 \usepackage{float}
 \usepackage{rotating}
 \usepackage{placeins}

\usepackage[numbers,sort&compress]{natbib}

\usepackage{tikz}
\usetikzlibrary{quantikz2}

\DeclarePairedDelimiter\ceil{\lceil}{\rceil}

\usepackage[list=false]{subcaption}
\usepackage{textgreek}
\usepackage{tikz-3dplot}
\usepackage[normalem]{ulem}
\usepackage[]{url}

\usetikzlibrary{decorations,decorations.pathreplacing,calc, arrows.meta,arrows,quantikz2}

\begin{document}

\title{Towards Quantum Machine Learning of Lattice Boltzmann Collision Operators for Fluid Dynamic Simulations}

\author{Wael Itani}
\affiliation{Tandon School of Engineering, New York University, New York, NY 11201, United States of America}
\affiliation{Maroun Semaan Faculty of Engineering \& Architecture, American University of Beirut, Beirut, Lebanon}
\email{wi07@aub.edu.lb}
\author{Katepalli R. Sreenivasan}
\affiliation{Tandon School of Engineering, New York University, New York, NY 11201, United States of America}
\email{katepalli.sreenivasan@nyu.edu}

\begin{abstract}
\title{We attempt the use of a unitary operator to approximate the lattice Boltzmann collision operator. We use a modified amplitude encoding to bypass the renormalization that would have required classical processing at every step (thus eroding any quantum advantage to be had). We describe the hard-wiring of the lattice Boltzmann symmetries into the quantum circuit and show that, for the specific case of the cavity flow, approximating the nonlinear system is limited to low velocities. These findings may help us understand better the possibilities of nonlinear simulations on a quantum computer, and also pave the way for a discussion on how quantum machine learning might be harnessed to address more complex problems.}
\end{abstract}

\section{Introduction}
\subsection{Lattice Boltzmann Method}
\subsubsection{Overview}
The lattice Boltzmann method (LBM) is concerned with solving for the probability distribution density of particles of a fluid at the mesoscale. It is characterized by an operator splitting of the Boltzmann equation, resulting in performing streaming and collision steps successively in each timestep, rather than in a coupled fashion as with finite-volume Navier-Stokes (N-S) solvers. This has advantages over the solution of the nonlinear advection term of N-S, which typically requires several iterations at every timestep \cite{wangGPUImplementedLatticeBoltzmann2024}. 

A closed form of the Boltzmann collision is obtained by invoking the BGK assumption, and writing the collision as an explicit relaxation to equilibrium. In the standard collision model, which uses the same assumption, the individual discrete densities are each allowed to relax to a Maxwellian equilibrium while preserving two of their moments, which are the hydrodynamic density and fluid velocity.

The lattice Boltzmann (LB) formulation uses neither the continuum-scale velocity fields addressed by Navier-Stokes solvers nor the microscopic description of molecular dynamics, but discrete probability densities $f$ of fluid parcels (or particles) moving on a $D$-dimensional lattice in $Q$ discrete directions. This makes it inherently suitable for use as the basis of quantum algorithms. This convenience stems from restricting the Boltzmann kinetic equation to a lattice, so that the velocity $\vec{c}_i$ of particles in the direction of the lattice vector $\vec{e}_i$ can be computed in a single time. Here $\vec{c}_i=c_i\vec{e}_i$. 

One may regard the lattice Boltzmann method for simulation of fluid flows either as a computational model based on the kinetic theory, or as an attempt to take lattice automata to the next level of sophistication. The historical context favors the latter view. The formulation that has gained popularity employs a truncated version of the equilibrium function and utilizes the instruction set available on computer processors. The advent of graphical processing units (GPU) has propelled the method into a brighter spotlight as its operations---collisions being local and nonlinear while streaming is nonlocal and linear---are amenable to massive parallelization. As quantum computers usher a new era of computing, we are persuaded to reconsider the approximations made in porting the method to this new hardware.

\subsection{Formulation}
\label{sec:form}

The Boltzmann equation
\begin{equation}
\label{BolEq}
    \frac{d{f}}{dt} = \frac{\partial{{f}}}{\partial{t}}+\vec{v}\cdot \vec{\nabla} {f} = {\Omega}
\end{equation}
 describes the probability density $f$ of the fluid in the position-momentum space, driven by advection due to continuum particle velocity $\vec{v}$ and collision $\Omega$ across space spanned by $\vec{x}$, and time $t$.  To arrive at the lattice Boltzmann formulation, the probability density $f$ from Eq.~(\ref{BolEq}) is discretized into $Q$ density distributions, each describing the fraction of fictitious fluid particles: moving in a given $D$-dimensional lattice, with $\vec{v}$ restricted to speeds $\vec{c}_i$ %$\vec{c}_i=c_i\vec{e}_i$, 
 defined in the directions of the lattice vectors $\vec{e}_i$. 

\begin{figure}[ht]
\centering
\captionsetup[subfigure]{justification=centering}

\subcaptionbox[D3Q27]{D3Q27\label{subfig:d3q27}}
[3cm]
{\resizebox{3cm}{!}{% set view angle
\tdplotsetmaincoords{74}{113}% rot x rot z
         \begin{tikzpicture}[tdplot_main_coords,
axis/.style={thick, ->, >=stealth'}]

\coordinate (O) at (0.5,0.5,0.5);

\def \a {1}       % unit cells along a (only integers!)
\def \b {1}       % unit cells along b (only integers!)
\def \c {1}       % unit cells along c (only integers!)

% lattice directions parallel to c
 \foreach \u in {0,1,...,\a}
    \foreach \v in {0,1,...,\b}
      \foreach \w in {0,1,...,\c}
        \draw[very thin,gray] (\u,\v,0) -- (\u,\v,\w); 
% lattice directions parallel to b
 \foreach \u in {0,1,...,\a}
    \foreach \v in {0,1,...,\b}
      \foreach \w in {0,1,...,\c}
        \draw[very thin,gray] (\u,0,\w) -- (\u,\v,\w); 
% lattice directions parallel to a
 \foreach \u in {0,1,...,\a}
    \foreach \v in {0,1,...,\b}
      \foreach \w in {0,1,...,\c}
        \draw[very thin, gray] (0,\v,\w) -- (\u,\v,\w);

% basis vectors
\draw plot [mark=*, mark size=1] coordinates{(O)};

% lattice directions to corners
 \foreach \u in {0,1,...,\a}
    \foreach \v in {0,1,...,\b}
      \foreach \w in {0,1,...,\c}
        \draw[thin,-latex,black](O) -- (\u,\v,\w);
        
% lattice directions to corners in 2D planes
% \foreach \u in {0,1,...,\a}
    \foreach \v in {0,1,...,\b}
      \foreach \w in {0,1,...,\c}
        \draw[thin,-latex,black](O) -- (0.5,\v,\w);
 \foreach \u in {0,1,...,\a}
   % \foreach \v in {0,1,...,\b}
      \foreach \w in {0,1,...,\c}
        \draw[thin,-latex,black](O) -- (\u,0.5,\w);
 \foreach \u in {0,1,...,\a}
    \foreach \v in {0,1,...,\b}
     % \foreach \w in {0,1,...,\c}
        \draw[thin,-latex,black](O) -- (\u,\v,0.5);

%Lattice vectors to edges in 2D planes
 \foreach \u in {0,1,...,\a}
    %\foreach \v in {0,1,...,\b}
     % \foreach \w in {0,1,...,\c}
        \draw[thin,-latex,black](O) -- (\u,0.5,0.5);
%\foreach \u in {0,1,...,\a}
    \foreach \v in {0,1,...,\b}
     % \foreach \w in {0,1,...,\c}
        \draw[thin,-latex,black](O) -- (0.5,\v,0.5); 
%\foreach \u in {0,1,...,\a}
    %\foreach \v in {0,1,...,\b}
      \foreach \w in {0,1,...,\c}
        \draw[thin,-latex,black](O) -- (0.5,0.5,\w); 

\end{tikzpicture}}}
\hspace{0.1\textwidth}
\subcaptionbox[D2Q9]{D2Q9\label{subfig:d2q9}}
[3cm]
{\resizebox{3cm}{!}{% set view angle
\tdplotsetmaincoords{74}{113}% rot x rot z
         \begin{tikzpicture}[tdplot_main_coords,
axis/.style={thick, ->, >=stealth'}]

\coordinate (O) at (0.5,0.5,0.5);

\def \a {1}       % unit cells along a (only integers!)
\def \b {1}       % unit cells along b (only integers!)
\def \c {1}       % unit cells along c (only integers!)

% lattice directions parallel to c
 \foreach \u in {0,1,...,\a}
    \foreach \v in {0,1,...,\b}
      \foreach \w in {0,1,...,\c}
        \draw[very thin,gray] (\u,\v,0) -- (\u,\v,\w); 
% lattice directions parallel to b
 \foreach \u in {0,1,...,\a}
    \foreach \v in {0,1,...,\b}
      \foreach \w in {0,1,...,\c}
        \draw[very thin,gray] (\u,0,\w) -- (\u,\v,\w); 
% lattice directions parallel to a
 \foreach \u in {0,1,...,\a}
    \foreach \v in {0,1,...,\b}
      \foreach \w in {0,1,...,\c}
        \draw[very thin, gray] (0,\v,\w) -- (\u,\v,\w);

% basis vectors
\draw plot [mark=*, mark size=1] coordinates{(O)};

% lattice directions to corners in 2D planes
% \foreach \u in {0,1,...,\a}
    \foreach \v in {0,1,...,\b}
      \foreach \w in {0,1,...,\c}
        \draw[thin,-latex,black](O) -- (0.5,\v,\w);

%Lattice vectors to edges in 2D planes
%\foreach \u in {0,1,...,\a}
    \foreach \v in {0,1,...,\b}
     % \foreach \w in {0,1,...,\c}
        \draw[thin,-latex,black](O) -- (0.5,\v,0.5); 
%\foreach \u in {0,1,...,\a}
    %\foreach \v in {0,1,...,\b}
      \foreach \w in {0,1,...,\c}
        \draw[thin,-latex,black](O) -- (0.5,0.5,\w); 

\end{tikzpicture}

% set view angle
\tdplotsetmaincoords{74}{113}% rot x rot z
         \begin{tikzpicture}[tdplot_main_coords,
axis/.style={thick, ->, >=stealth'}]

\coordinate (O) at (0.5,0.5,0.5);

\def \a {1}       % unit cells along a (only integers!)
\def \b {1}       % unit cells along b (only integers!)
\def \c {1}       % unit cells along c (only integers!)

% lattice directions parallel to c
 \foreach \v in {0,1,...,\b}
      \foreach \w in {0,1,...,\c}
        \draw[very thin,gray] (0.5,\v,0) -- (0.5,\v,\w); 
% lattice directions parallel to b
 \foreach \v in {0,1,...,\b}
      \foreach \w in {0,1,...,\c}
        \draw[very thin,gray] (0.5,0,\w) -- (0.5,\v,\w);

% basis vectors
\draw plot [mark=*, mark size=1] coordinates{(O)};

% lattice directions to corners in 2D planes
% \foreach \u in {0,1,...,\a}
    \foreach \v in {0,1,...,\b}
      \foreach \w in {0,1,...,\c}
        \draw[thin,-latex,black](O) -- (0.5,\v,\w);

%Lattice vectors to edges in 2D planes
%\foreach \u in {0,1,...,\a}
    \foreach \v in {0,1,...,\b}
     % \foreach \w in {0,1,...,\c}
        \draw[thin,-latex,black](O) -- (0.5,\v,0.5); 
%\foreach \u in {0,1,...,\a}
    %\foreach \v in {0,1,...,\b}
      \foreach \w in {0,1,...,\c}
        \draw[thin,-latex,black](O) -- (0.5,0.5,\w); 

\end{tikzpicture}}}
\hspace{0.1\textwidth}
\subcaptionbox[D1Q3]{D1Q3\label{subfig:d1q3}}
[3cm]
{\resizebox{3cm}{!}{% set view angle
\tdplotsetmaincoords{74}{113}% rot x rot z
         \begin{tikzpicture}[tdplot_main_coords,
axis/.style={thick, ->, >=stealth'}]

\coordinate (O) at (0.5,0.5,0.5);

\def \a {1}       % unit cells along a (only integers!)
\def \b {1}       % unit cells along b (only integers!)
\def \c {1}       % unit cells along c (only integers!)

% lattice directions parallel to c
 \foreach \u in {0,1,...,\a}
    \foreach \v in {0,1,...,\b}
      \foreach \w in {0,1,...,\c}
        \draw[very thin,gray] (\u,\v,0) -- (\u,\v,\w); 
% lattice directions parallel to b
 \foreach \u in {0,1,...,\a}
    \foreach \v in {0,1,...,\b}
      \foreach \w in {0,1,...,\c}
        \draw[very thin,gray] (\u,0,\w) -- (\u,\v,\w); 
% lattice directions parallel to a
 \foreach \u in {0,1,...,\a}
    \foreach \v in {0,1,...,\b}
      \foreach \w in {0,1,...,\c}
        \draw[very thin, gray] (0,\v,\w) -- (\u,\v,\w);

% basis vectors
\draw plot [mark=*, mark size=1] coordinates{(O)};

%Lattice vectors to edges in 2D planes
%\foreach \u in {0,1,...,\a}
    \foreach \v in {0,1,...,\b}
     % \foreach \w in {0,1,...,\c}
        \draw[thin,-latex,black](O) -- (0.5,\v,0.5); 
\end{tikzpicture}

% set view angle
\tdplotsetmaincoords{74}{113}% rot x rot z
         \begin{tikzpicture}[tdplot_main_coords,
axis/.style={thick, ->, >=stealth'}]

\coordinate (O) at (0.5,0.5,0.5);

\def \a {1}       % unit cells along a (only integers!)
\def \b {1}       % unit cells along b (only integers!)
\def \c {1}       % unit cells along c (only integers!)

% lattice directions parallel to c
 \foreach \u in {0,1,...,\a}
    \foreach \v in {0,1,...,\b}
      \foreach \w in {0,1,...,\c}
        \draw[very thin,white] (\u,\v,0) -- (\u,\v,\w); 
% lattice directions parallel to b
 \foreach \u in {0,1,...,\a}
    \foreach \v in {0,1,...,\b}
      \foreach \w in {0,1,...,\c}
        \draw[very thin,white] (\u,0,\w) -- (\u,\v,\w); 
% lattice directions parallel to a
 \foreach \u in {0,1,...,\a}
    \foreach \v in {0,1,...,\b}
      \foreach \w in {0,1,...,\c}
        \draw[very thin,white] (0,\v,\w) -- (\u,\v,\w);

% basis vectors
\draw plot [mark=*, mark size=1] coordinates{(O)};

%Lattice vectors to edges in 2D planes
%\foreach \u in {0,1,...,\a}
    \foreach \v in {0,1,...,\b}
     % \foreach \w in {0,1,...,\c}
        \draw[thin,-latex,black](O) -- (0.5,\v,0.5); 
\end{tikzpicture}}}
\caption{Different lattice configurations in three, two and one dimensions}
\label{fig:lattices}
\end{figure}
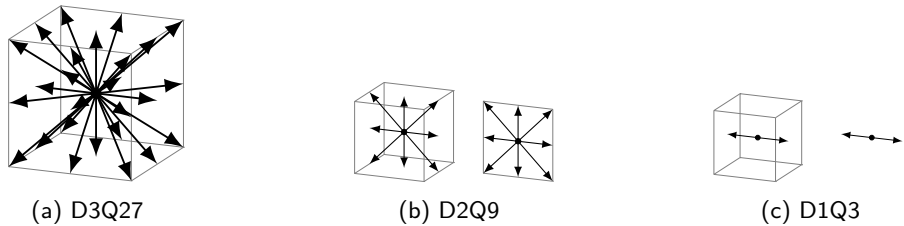

The discretized lattice Botlzmann equation under the BGK approximation takes the form
\begin{equation}
\label{lBolEq}
    \frac{1}{\Delta t}(f_i(\vec{x}+\vec{c}_i\Delta t, t + \Delta t) - f_i(\vec{x},t)) = - \frac{1}{\tau}(f_i(\vec{x},t)-f_i^{eq}(\vec{x},t))
\end{equation}
for the simple case of a single-phase fluid (whose components are permeable to each other and can be considered well-mixed and homogeneous \cite{billLatticeBoltzmannMethod}.) The discrete probability density is the fundamental variable of the lattice Boltzmann approach. It describes the probability of finding a fluid particle at a given point defined by the position vector $\vec{x}$, at a given instant of time $t$, with a particular speed \cite{randlesPerformanceAnalysisLattice2013}. The lattice Boltzmann proceeds by updating the discrete probability densities at each cell in two steps: collision and advection/streaming. The evolution of the solutions involves these two operations at each time step. A quantum algorithm for lattice Boltzmann, then, requires an implementation of the two steps: collision and streaming.

\subsubsection{Streaming}
Streaming proceeds by propagating each discrete density along its respective lattice direction to the neighboring cell. As already stated, streaming is nonlocal and linear, exact in reference to the round-off error when performed on a classical computer, as
\begin{equation}
\label{strBolEq}
    f_i(\vec{x},t + \Delta t) \xrightarrow{} f_i(\vec{x}+\vec{c}_i\Delta t, t + \Delta t).
\end{equation}
\subsubsection{Collision}
Collision is subsumed as a relaxation towards equilibrium, a nonlinear operation dependent on terms local to each cell. Under the BGK approximation, the collision term $\Omega_i$ is simply
\begin{equation}
    \Omega_i = -\frac{1}{\tau} (f_i -f_i^{eq}).
\end{equation}

During collision, the discrete densities are updated based on the local densities in the same cell according to the collision term, which is on the right hand side of the discretized equation, Eq.~(\ref{lBolEq}), so that
\begin{equation}
\label{colBolEq}
    f_i(\vec{x}, t + \Delta t) = f_i(\vec{x},t) +\Delta t\Omega_i(\vec{f}).
\end{equation}

\subsubsection{Equilibrium Distribution for a Single-Phase Fluid}
From the right hand side of Eq.~(\ref{lBolEq}), one can see that the collision process is a relaxation towards an equilibrium distribution.  The equilibrium distribution $f^{eq}_i$ is written as a function of the lattice speed of sound $c_s$, the fluid density $\rho$, the lattice vectors $\vec{e}_i$, and the flow velocity $\vec{u}$, where
\begin{equation}
    \rho = {\Sigma}_{i=1}^Q f_i,
\end{equation}
\begin{equation}
    \vec{u}=\frac{c}{\rho} {\Sigma}_{i=1}^Q f_i\vec{e}_i,
\end{equation}
and
\begin{equation}
    \Sigma_{i=1}^Q w_i = 1
\end{equation}
A common model for the equilibrium function for a single-phase fluid is
\begin{equation}
    f_i^{eq}(\vec{x},t) = w_i\rho(1+\frac{1}{c_s^2}\vec{e}_i\cdot\vec{u}+\frac{1}{2}((\frac{1}{c_s^2}\vec{e}_i\cdot \vec{u})^2-\frac{1}{c_s^2}\vec{u}\cdot\vec{u})).
\end{equation}
 This equilibrium function is, in fact, an approximation of the Boltzmann distribution, a Gaussian in local Mach number $Ma$, up to $O(Ma^2)$, rendering the lattice Boltzmann method in this formulation, and any linearization thereof, subject to the assumption $Ma<<1$. The nonlinearity is also more directly linked to the Mach number, as correctly concluded in \cite{liPotentialQuantumAdvantage2024} since the Boltzmann distribution is {nonlinear} in Mach number. This is contrary to the Navier-Stokes equations whose nonlinearity is controlled by the Reynolds number. This is not to say that Reynolds number does not play a role here, as it is proportional to the relaxation frequency and controls the lattice size.
 
We note that the lattice vectors $\vec{e}_i$ do not correspond to unit vectors, and are generally not orthonormal. In a single dimension, the left and right vectors have nonzero inner product of $-1$. In higher dimensions, we can identify such a pair for each direction, in addition to nonzero inner products with and between diagonal lattice vectors shown in figure \ref{fig:lattices}.

\subsection{Prior Attempts at Quantum Algorithms for Lattice Boltzmann}
The evolution of quantum algorithms in this direction has largely followed that of their classical counterparts, with the lattice Boltzmann algorithms arriving chronologically after lattice gas algorithms \cite{yepezLatticeGasQuantumComputation1998}. Recently, quantum lattice gas algorithms have been overhauled in \cite{kocherlaFullyQuantumAlgorithm2024}. This work translates that of  \cite{yepezQuantumLatticeGasModel2002}, which used the now-superseded Type-II quantum computers, to the modern framework of universal, gate-based, quantum computing. Reference \cite{kocherlaFullyQuantumAlgorithm2024} also demonstrates an algorithm for the diffusion equation. However, the number of qubits needed in this algorithm scales linearly with the grid points. Solving the diffusion equation in 1D for 50 sites required them to use 102 qubits. This is in line with the suggestion of \cite{schalkersImportanceDataEncoding2024} not to encode the values of lattice variables at different lattice sites in superposition on a quantum computer. However, this means starting from the premise of giving up on any quantum advantage in the hardware resources. 

On the other end of the spectrum, we have algorithms that compromise the promise of quantum computational complexity advantage. In \cite{ljubomirQuantumAlgorithmNavier2022, kocherlaTwocircuitApproachReducing2024}, the algorithm implicitly relies on measuring the solution at every step of evolution such that their quantum computational complexity ends up scaling at least linearly in the number of grid points (lattice sites). We note that these two approaches are similar and that one could, in fact, improve upon either the qubit complexity of the former, or the gate complexity of the latter, by measuring every so often, or by running several copies of the circuit to reduce the measurement frequency, respectively. 

A third approach that complements this line of work is to encode lattice site positions in superposition in a quantum register in the computational basis, as is standard for linear problems, and as we have done in \cite{itaniQuantumAlgorithmLattice2024}. However, rather than a single copy, several copies of the register are prepared. Discrete densities are then streamed in all discrete density-lattice basis combinations, after which amplitude amplification is used to select the substate with the correct discrete-density lattice direction. The exact number of copies that need to be prepared would scale polynomially in the number of dimensions. This is on par with the qubit scaling achieved by linearizing the operator classically and including higher-order cross-term variables from neighboring cells, while writing the matrix for the quantum linear system algorithm (QLSA) \cite{liPotentialQuantumAdvantage2024}.

\begin{figure*}[!htpb]
    \centering
    \includegraphics[width=\columnwidth]{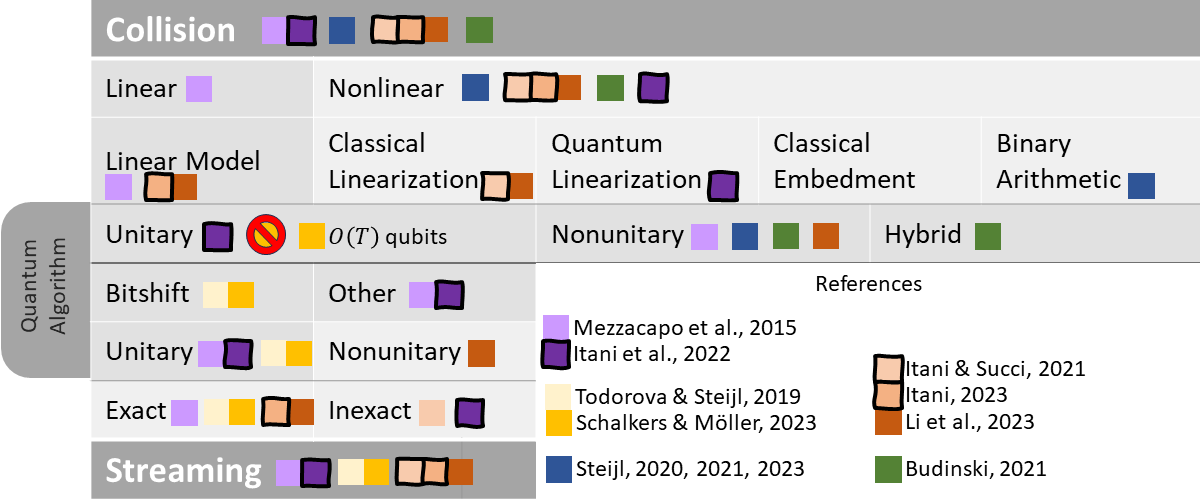}
    \caption{Main lines of research for the approaches on a quantum algorithm for the lattice Boltzmann method.}
    \label{fig:lit-rev}
\end{figure*}

We organize published attempts of quantum algorithms for lattice Boltzmann simulations in Fig.~\ref{fig:lit-rev}. The relevant references are \cite{mezzacapoQuantumSimulatorTransport2015,itaniQuantumAlgorithmLattice2024,todorovaQuantumAlgorithmCollisionless2020a,schalkersEfficientFailsafeQuantum2024,steijlParallelEvaluationQuantum2018a,steijlQuantumAlgorithmsNonlinear2020,steijlQuantumCircuitImplementation2023,itaniAnalysisCarlemanLinearization2022,itaniLatticeBoltzmannLinear2023a,liPotentialQuantumAdvantage2024,budinskiQuantumAlgorithmAdvection2021}. The more recent works, \cite{bakkerQuantumCarlemanLinearization2024}, \cite{ljubomirQuantumAlgorithmNavier2022,kocherlaTwocircuitApproachReducing2024}, \cite{budinskiEfficientParallelizationQuantum2023}, and \cite{kocherlaFullyQuantumAlgorithm2024}, fall under \cite{liPotentialQuantumAdvantage2024}, \cite{budinskiQuantumAlgorithmAdvection2021}, \cite{todorovaQuantumAlgorithmCollisionless2020a}, and \cite{schalkersImportanceDataEncoding2024}, respectively. Other recent literature that does not use the lattice methods, such as \cite{ingelmannTwoQuantumAlgorithms2024,au-yeungQuantumAlgorithmSmoothed2024,gaitanCircuitImplementationOracles2024,pfefferReducedorderModelingTwodimensional2023}, as well as \cite{ameriQuantumAlgorithmLinear2023}, which deals with the Vlasov equation (equivalent to Boltzmann equation with force term with a linear collision model), have been omitted.

We have filled the gap of considering inexact streaming in \cite{itaniQuantumAlgorithmLattice2024}. Inexact streaming is incompatible with the state-of-the-art implementation of the streaming step \cite{schalkersEfficientFailsafeQuantum2024,todorovaQuantumAlgorithmCollisionless2020a}, which requires the discrete densities to be encoded in superposition in a quantum register, including the collision term that we introduced in \cite{itaniQuantumAlgorithmLattice2024}.

We have shown in \cite{itaniAnalysisCarlemanLinearization2022} that the quadratic (nonlinear) BGK-type collision term maps to a finite linear system through a Carleman linearization of second order. However, the linearized system gives rise to variables mapping from coupled, second-order variables in the original formulation. In \cite{itaniAnalysisCarlemanLinearization2022}, we conclude that this further exacerbates the demanding memory requirements of the lattice Boltzmann method on classical computers, but makes it adequate for treatment on quantum computers, as has been done by others \cite{liuEfficientQuantumAlgorithm2021,liPotentialQuantumAdvantage2024}. Rather than using a predefined linearization, we seek to enable the search for a linear operator that approximates the collision operator. Rather than hard-wiring a predetermined basis, as with the Carleman linearization and linear embedding approaches, by enconding the discrete density values and lattice vector indices on separate registers, as explained in Sec.~\ref{sec:qmlmethods}, we allow the search for a basis transform by including unitaries that act solely on the lattice index register. Since the unitaries considered constitute linear operators acting on the circuit, with their nonlinearity only appearing when considering their action on the decoded values (i.e. understanding the action of the controlled-gates on the encoded lattice indices and values), we are approximating the nonlinear collision operator in the real space with a linear one in the finite Hilbert space. Rather than being predetermined, this ``linearization" is dependent on the dataset considered. We thus see in Sec.~\ref{sec:lowbound} that an appropriate range of applicability must be determined for the collision operator. Recent work has presented an alternative, more systematic way, to hard-wire the symmetries into the construction of the collision circuit \cite{lacatusSurrogateQuantumCircuit2025}.

The separate linear streaming and nonlinear collision steps of lattice Boltzmann also make it an adequate target for a function-fitting approach. The search for a collision operator is then cast as a regression problem. We would only have to train the network to function fit the collision model, which then becomes transferable to other flow configurations with the same underlying physics.

In contrast to \cite{itaniQuantumAlgorithmLattice2024}, we maintain unitary exact streaming by sacrificing the asymptotic exactness of the collision. This is done by empirically constructing an approximate collision operator through quantum machine learning. We assess the performance of this construction.

\subsection{Brief Overview of Quantum Machine Learning}
 Machine learning is typically divided into supervised, unsupervised and reinforcement learning (see, e.g., Ref.~\cite{pandey_perspective_2020}). Whereas in the first a computer is provided with a dataset of input-output pairs, in the second only an ``output" is provided, from which the computer is expected to make sense, typically to cluster and reduce its dimensionality, without regard to where the data has been mapped from. In reinforcement learning, neither the input nor the output is provided, but an action space is provided to the computer as an agent for evaluating the actions based on set of rules provided, typically a cost function. In what follows, any reference to machine learning is made to the supervised case in which the training is done against a dataset of inputs-outputs.

Artificial neural networks are n-dimensional graphs where the nodes share weighed connections. The graph is traversed directionally as to update the value of a given node based on the weighed value of previously traversed nodes. Classically, there is no restriction on traversing a given node more than once, as is clearly done by ``feedback networks". A ``feedforward" network by comparison is traversed in layers, such that the values in each layer are updated before the next one, closer to the output. The activation function responsible for determining the node values is typically the same, and the connections between the nodes are a hyper-parameter of the network, designed {\it a priori} to training. The training process then consists of optimizing the weights so that the nodes designated as outputs yield values as close as possible to the true values, for given input values at the nodes designated as input.

Quantum machine learning as a term initially referred to the utilization of quantum computers to perform (and accelerate) tasks within classical machine learning \cite{schuldInnovatingMachineLearning}. Once quantum circuits were accepted for what they are, quantum machine learning has come to indicate the ``training" of quantum circuit parameters, by invoking a classical optimization step for a specified cost function. This special case of variational algorithms yielded the term ``quantum neural network" (QNN) for parameterized quantum circuits. Indeed, quantum circuits constructed variationally to minimize a cost-function measured over a dataset form linear, symmetric neural networks. 

Translating nonlinear activation functions, proven essential to the performance of classical artificial neural networks, simply stated as neural networks, to a quantum mechanical framework is an open question. Quantum neural networks models have been proposed since the 1995 \cite{dendukuriDefiningQuantumNeural2020}, and have taken 20 years to be formalized \cite{schuldIntroductionQuantumMachine2015}, after which targeted efforts have been made to transport the nonlinear activation functions characteristic of classical machine learning to the quantum framework. The approaches to achieving the latter differ between approximating the classical implementation \cite{inajetovicEnablingNonLinearQuantum2023} and using inherently quantum processes like measurement \cite{giliIntroducingNonlinearActivations2023}. 

Choosing an adequate variational quantum circuit for a given quantum machine learning task also remains an open challenge, especially in regards to avoiding a barren plateau and ensuring trainability \cite{kruseVariationalQuantumCircuit2024}. Equally important is the data encoding strategy used \cite{schuldEffectDataEncoding2021}. The expressivity of the variational model and its domain of applicability are inherently linked to the choice of circuit architecture and encoding \cite{schuldInnovatingMachineLearning}.

Machine learning is not closed under quantum computing. The backpropagation step needed to ``learn" and optimize the weights is done classically. One advantage of quantum machine learning is that the gradients for the weights could be computed analytically for use in the optimization procedure \cite{schuldEvaluatingAnalyticGradients2019}. The rotation gates used in a quantum circuit have theoretically continuous parameters that are adjusted in the training process. Essentially, the circuit output is sampled from a probability distribution, and, with repeated executions and measurements, an expectation value is approached asymptotically. This expectation value varies smoothly with gate parameters, so as to allow a gradient-descent based optimization approach. The parameter-shift rule dictates that the gradient of expectation values obtained from a quantum circuit could be obtained by two executions of the circuit itself with one parameter varied between the two executions. Whereas the rule was first derived for unitaries corresponding to Pauli Hamiltonians, at the cost of an additional ancilla qubit, it could be generalized to any gate with a linear combination of unitaries routine \cite{schuldEvaluatingAnalyticGradients2019}.

\section{Methods}
\label{sec:qmlmethods}
To move from approximations of a well-defined nonlinear function, the collision term of the lattice Boltzmann formulation, to a well-defined quantum machine learning problem, we must first obtain a suitable dataset for the training. Choices have to be made regarding the data encoding in the quantum circuit, as well as the ansatz circuit architecture.

\subsection{Dataset Generation}
We make use of the data generation methodology of \cite{corbettaLearningLatticeBoltzmann2023}. 
\begin{enumerate}
\setcounter{enumi}{0}
    \item A maximum velocity magnitude $u_{max}$, an allowed range of fluctuation for the density $[\rho_{min}, \rho_{max}]$, standard deviation $[\sigma_{min},\sigma_{max}]$, and relaxation constant $[\tau_{min},\tau_{max}]$ and dataset size $N$ are specified.
    \item $\rho$ is sampled from a uniform distribution $U(\rho_{min},\rho_{max})$.
    \item The velocity magnitude $u_0$ and direction specified by angle $\theta$ are sampled form $U(0,u_{max})$ and $U(0,2\pi)$, respectively, giving $\vec u = u_0(\cos(\theta), \sin(\theta))^T$.
    \item The corresponding equilibrium distribution is calculated $f_i^{eq}(\rho, \vec u)$.
    \item A standard deviation $\sigma$ is sampled from a uniform distribution $U(\sigma_{min},\sigma_{max})$.
    \item Random discrete density perturbations $f_i^{'neq}$ are sampled from a normal distribution $N(0,\sigma^2)$ centered around zero with the sampled standard deviation.
    \item Not to alter the prescribed hydrodynamic variables $\rho$ and $\vec u$, the moments of the perturbation are computed and used to shift the distribution as $f^{neq}_i = f_i^{'neq} - \frac{1}{9} \rho' - \frac{1}{6} \vec{e}_i \cdot \rho' \vec u'$ .
    \item The pre-collision discrete density is then specified as the sum of the equilibrium and non-equilibrium components $f^{pre}_i = f_i^{eq}+f_i^{neq}$.
    \item A relaxation constant $\tau$ is sampled from a uniform distribution $U(\tau_{min},\tau_{\max})$.
    \item The post-collision distribution is then analytically evaluated as $f_i^{post} = f_i^{pre}-\frac{\Delta t}{\tau}(f_i^{pre}-f_i^{eq})$
    \item The equilibrium, pre-collision, and post-collision discrete densities, $\vec f^{eq}$, $\vec f^{pre}$, and $\vec f^{post}$ are checked for positivity, and the whole data point is discarded if a negative entry is found.
    \item Steps $2-11$ are repeated as necessary to build a dataset of prescribed size $N$.
\end{enumerate}

For our case, the values for the prescribed quantities used are presented in Tab.~\ref{tab:datagen}.

\begin{table}
    \centering
    \begin{tabular}{|c|c|l|} \hline 
         Quantity& Minimum Value &Maximum Value\\ \hline 
         $u$& 0&0.1\\ 
         $\rho$&  0.95&1.05\\
         $\sigma$&  $10^{-15}$&$5\times 10^{-4}$\\
 $\tau$& 1&1\\ \hline
    \end{tabular}
    \caption[Summary of parameters used in generating a dataset of random pre- and post-collision values]{Summary of the prescribed quantities describing the distributions of random variables used in generating the dataset}
    \label{tab:datagen}
\end{table}

\subsection{Lattice Basis Vector Encoding} 
For comparison with \cite{corbettaLearningLatticeBoltzmann2023}, we limit ourselves to the two-dimensional case. Moreover, a lattice with diagonal discrete velocities is required to maintain the lattice Boltzmann advantage of recovering second-order statistics with a second-order method. As discussed \cite{itaniQuantumAlgorithmLattice2024}, to be able to perform streaming, we require the discrete densities to be encoded in superposition. The simplest way to do so would be to use the computational basis
\begin{equation}
    \frac{1}{\sqrt{Q}}\Sigma_{i=0}^{Q-1}\ket{i}\ket{f_i}.
\end{equation}
However, at the cost of using $2D$ qubits instead of $\ceil{\log_2{(Q)}} = \ceil{\log_2{(3^D)}} = \ceil{D\log_2{(3)}}$ qubits, we encode the components of the lattice basis vector, while simplifying the construction of a symmetry-preserving neural network down the road. Two of the $2D$ qubits are used to encode the component of the lattice basis vector in a given dimension, as detailed in Tab.~\ref{tab:ei}.
\begin{table}[!htbp]
        \caption{States encoding the different directions for a given dimension}
\begin{tabularx}{\textwidth}{|X|X|}
  \hline
  Direction & $\ket{\vec{e}_i,d} = \ket{e_{i,0} e_{i,1}}$
  \\\hline
Stationary & $\frac{1}{\sqrt{2}}(\ket{01}+\ket{10})$
\\ Positive Axis & $\ket{11}$
\\Negative Axis & $\ket{00}$
\\\hline
  \end{tabularx}
  \label{tab:ei}
\end{table}
Therefore, the overall state of the system encoding all $Q$ discrete densities at a single lattice site is
\begin{equation}
    \frac{1}{\sqrt{Q}}\Sigma_{i=1}^Q \ket{\vec{e}_i}\ket{f_i},
\end{equation}
to which we refer as the 2D basis. It is important to note that the said basis is convenient for the streaming operations when the quantum circuit is being evaluated or used. However, the computational basis remains preferred when the circuit is being trained, as it is compatible with a unitary implementation of the symmetry transforms discussed below. On the other hand, due to the definition of the stationary state in superposition, the 2D basis would need to use the full, $16$, storage capacity of the $4$ qubit by storing four copies of the rest particle discrete density, and two copies of each discrete density along a cardinal directional. This duplicity needed to achieve the transforms while maintaining consistent encoding renders the implementation thereof irreversible, and thus non-unitary. As explained below, the reversibility is essential for making use of the symmetries in the training.

\subsection{Modified Amplitude Encoding of Discrete Densities}
The standard amplitude encoding technique usually envolves renormalizing the data such that the $L2$ is unity when they are encoded in superposition into $n$ qubits, as
\begin{equation}
    \ket{\vec{y}} = \frac{1}{\sqrt{\norm{\vec{y}}_2}}\Sigma_{i=0}^{N-1} y_i\ket{i}
\end{equation}
for a vector of length $N= 2^n$. However, at the cost of an additional qubit, one may chose to put the index register in equal superposition,
\begin{equation}
    \ket{\vec{y}} = \frac{1}{\sqrt{N}}\Sigma_{i=0}^{N-1}\ket{i}\ket{0},
\end{equation}
and to encode the values with a constant normalization factor, as
\begin{equation}
    \ket{\vec{y}} = \frac{1}{\sqrt{N}}\Sigma_{i=0}^{N-1}\ket{i}(\sqrt{1-y_i^2}\ket{0}+y_i\ket{1}).
\end{equation}
For the case of lattice Boltzmann discrete density probabilities, they are bound from above by unity, such that there is no need to worry about rescaling. That the probabilities are positive definite also simplifies the readout process. The modified version of amplitude encoding could be thought of as a subset of a generalized binary encoding in which state superposition is utilized. It is worthy to note that such a definition of the encoding enables us to use the streaming and boundary condition operators defined in \cite{todorovaQuantumAlgorithmCollisionless2020a} and used in \cite{itaniQuantumAlgorithmLattice2024} without modification, which achieves the exact streaming equivalent to the classical operation.

\subsection{Neural Network Training and Architecture}
\subsubsection{Training}
To perform the training, we must choose:
\begin{itemize}
    \item Sample size
    \item Batch size
    \item Number of epochs 
    \item Learning rate
    \item Optimizer
    \item Loss function.
\end{itemize}
\paragraph{Samples and Batch Sizes}
The sample size is the overall count of samples seen by the network in one epoch. The batch size is the count of samples seen at once by the network. This is constrained by the memory available on the compute node. Each epoch consists of several batches such that, in all batches, the network has collectively seen all samples. We have
\begin{equation}
    Batches / Epoch = \frac{Sample\; Size}{Batch\;Size}
\end{equation}
Typically, the network sees all samples during each epoch before calculating the loss gradients and updating the circuit weights (parameters). This means that the sample size is equal to the size of the training dataset. However, in cases where the dataset is too large, the training may be set up to draw a random subset from the dataset at each epoch. Whether the training uses a sample size equal to the size of the training dataset---which means the network learns from the same samples at every epoch---or draws samples randomly (without replacement within an epoch), for every epoch, is indicated for each case considered.

\paragraph{Loss Function}
To calculate the loss, we use the standard root-mean-square error, $RSME$, which is a function of $N$ samples of a vector $\vec g$:
\begin{equation}
    Loss = RMSE(\vec g) =  \sqrt{\frac{1}{N}\Sigma_{n=1}^N(\Sigma_{\forall g_{i,n} \in \vec g_n} (\Tilde g_{i,n}-g_{i,n})^2)}.
\end{equation}
Here $g$ denotes the true value and $\Tilde{g}$ the predicted one. In addition to the discrete densities, we include the error of the hydrodynamic quantities conserved under collision, namely the density $\rho$ and the velocity norm $\norm{u}$, and note that the average is taken over the samples
\begin{equation}
    Loss = \sqrt{\frac{1}{N}\Sigma_{n=1}^N\frac{1}{3}((\vec{\Tilde{f}}_n-\vec f_n)^2+(\Tilde\rho_n-\rho_n)^2+(\Tilde{\norm{u}}_n-\norm{u}_n)^2)}.
\end{equation}

\subsection{Collision Ansatz Circuit Architecture}
We start by providing a high-level circuit diagram which shows how the collision ansatz works with the other components of the circuit that are necessary for training. We then explain the building blocks used in constructing the variations of the collision ansatz, and present each variation considered. In each of the cases considered, the quantum circuit has been implemented using the Pennylane Python library, and simulated using the $default.qubit$ device available in the library. 

As described in the section on encoding, two registers are necessary for the encoding of the discrete densities, a lattice basis vector $\ket{\vec{e}_i}$ register encoding its direction, and a separate register encoding its value $\ket{f_i}$. In addition to these registers, the algorithm requires $D$ registers to keep track of the lattice sites of the discrete densities and to perform streaming, which is not shown here. For the training, the ancilla qubits are required to perform the symmetry transforms. The ancilla qubits needed are $1$, $3$ and $6$, respectively for $D =1$, $2$ and $3$. The diagrams of the collision circuit ansatz are subsequently provided with only the discrete density (lattice basis vector, and value) registers displayed.
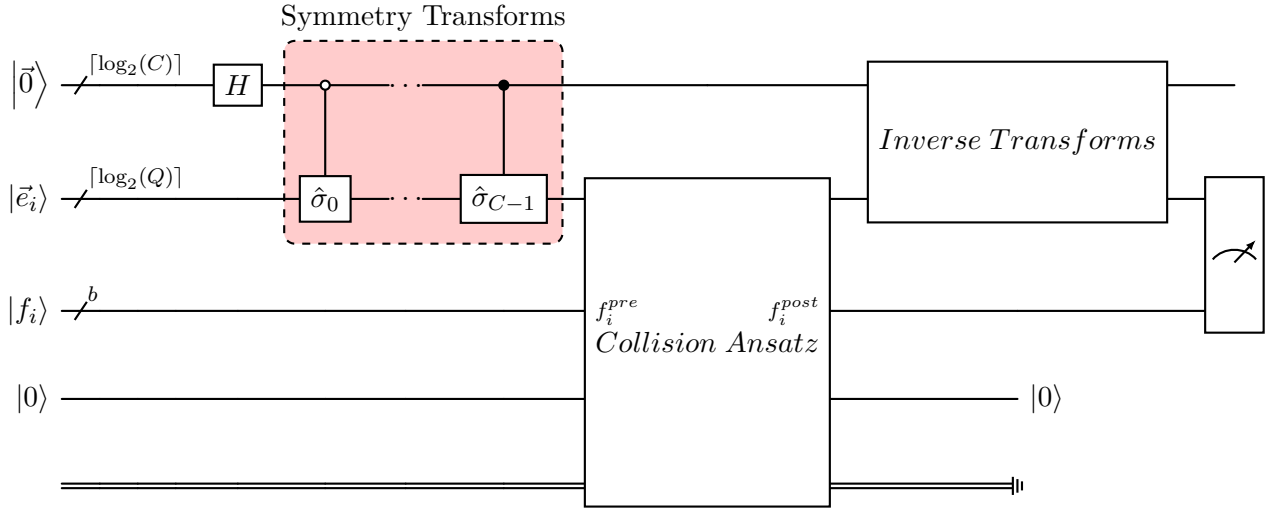
\begin{figure}[!htb]
    \centering
\begin{quantikz}[row sep = 25]
\lstick{$\ket{\vec{0}}$}&\qwbundle{\lceil \log_2 (C) \rceil}&&& \gate{H} &
\ctrl[open]{1}\gategroup[2,steps=3,style={dashed,rounded corners,fill=red!20, inner xsep=2pt},background,label style={label position=above,anchor=north,yshift=+0.4cm}]{Symmetry Transforms} &\ldots& \ctrl[]{1}&&\gate[2][1.5cm]{Inverse\; Transforms}& 
\\\lstick{$\ket{\vec{e}_i}$}&\qwbundle{\lceil \log_2(Q)\rceil}&&&& \gate{\text{$\hat{\sigma}_0$}} & \ldots&  \gate{\text{$\hat{\sigma}_{C-1}$}}& \gate[4][1.5cm]{Collision\;Ansatz}& &\meter[2]{}
\\\lstick{$\ket{f_i}$}&\qwbundle{b}&&&&&&&\gateinput{$f_i^{pre}$}\gateoutput{$f_i^{post}$}&&
\\\lstick{$\ket{0}$}&&&&&&&&&\rstick{$\ket{0}$}
\\&\setwiretype{c} &&&&&&&&\ground{}
\end{quantikz}

    \caption[Overview of the quantum circuit used for training]{Overview of the quantum circuit used for training, including the lattice basis vector, value, and lattice site registers, as well as the ancilla qubits.}
    \label{fig:general-circuit}
\end{figure}

 In addition to a controlled general rotation gate, controlled-$U3$, and a controlled rotation around the y-axis, controlled-$RY$, in the construction of our ansatz, we make extensive use of entangling layers, as explained below.
\subsubsection{Entangling Layer}
\label{subsec:el}
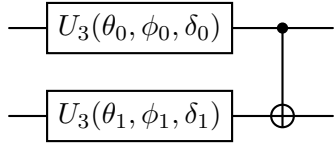
\begin{figure}
    \centering
    \begin{quantikz}
        &\gate{U_3(\theta_0,\phi_0,\delta_0)}&\ctrl[]{1}&\\
        &\gate{U_3(\theta_1,\phi_1,\delta_1)}&\targ{}&
    \end{quantikz}
    \caption{A quantum circuit showing the minimum viable entangling layer}
    \label{fig:ent-block}
\end{figure}

In all variations of the collision ansatz an entangling layer is used. The essential unit of an entangling layer is three-dimensional rotation $U3(\theta,\phi,\delta)$ gate applied to each qubit individually, followed by controlled-NOT (CNOT) gates applied to pairs of qubits. This building block, shown in Fig.~\ref{fig:ent-block}  for the cases of two qubits, is repeated several times with different parameters for the $U3$ gates. The ``layers" of an entangling layer refer to the number of repetitions of the essential unit applied consecutively with different parameters.
Among those parameters is the application of the CNOT gates. The distance between the control and target qubits is usually shifted by $+1$ with every repetition, such that the $n^{th}$ layer the qubits are $(n-1) [qc]$ qubits apart where $qc$ is the number of qubits to which the entangling layer is applied.

\begin{figure}
    \centering
    \begin{quantikz}
        &\gate{\text{$U_3(\theta_{0,l},\phi_{0,l},\delta_{0,l})$}}\gategroup[3,steps=4,style={solid,rounded corners,fill=red!0, inner xsep=2pt},background,label style={label position=above,anchor=north,yshift=+0.4cm}]{\text{{$\times\;l$}\;layers}}&\ctrl[]{1}&&\targ{}&\\
        &\gate{\text{$U_3(\theta_{1,l},\phi_{1,l},\delta_{1,l})$}}&\targ{}&\ctrl[]{1}&&\\
        &\gate[]{\text{$U_3(\theta_{2,l},\phi_{2,l},\delta_{2,l})$}}&&\targ{}&\ctrl[]{-2}&
    \end{quantikz}
    \caption{A quantum circuit showing an entangling layer with periodic CNOTs}
    \label{fig:ent-circ}
\end{figure}
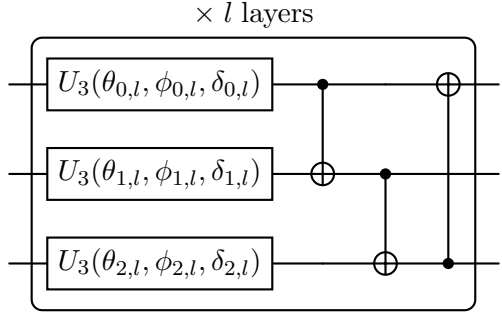

Another aspect the CNOT gates which must be chosen for construction is whether to consider linear or circular (periodic) boundaries. For simplicity, we discuss this when the two-qubit gate is applied to consecutive qubits. The CNOT gates could be applied in a linear fashion, as in Fig.~\ref{fig:ent-block}, such that they act upon each pair of consecutive qubits, and, except for the first and last qubits, each qubit is a target and a control for two different CNOTs exactly once. Alternatively, they could be applied in a periodic or circular fashion in which the first and last qubits are considered consecutive, as in Fig.~\ref{fig:ent-circ}.
The order of application of the CNOTs is a design choice. The application of the set of CNOTs more than once is equivalent to having more than one repetition of the essential unit, with only the first set of rotation parameters being non-zero. It is noteworthy that an entangling layer is the quantum equivalent of a classical convolution layer.

The parameters of the rotation gates could be constrained such that one can think of an essential layer with rotations around only one or two axes. When it is around a single axis, the entangling layer is known as a ``Basic Entangling Layer" (BEL), compared to the ``Strong Entangling Layer" (SEL) label for the general case. An example of a basic entangling layer would be one that restricts the rotations solely around the y-axis of the qubit, effectively replacing $U3(\theta,\phi,\delta)$ with $RY(\phi)$. In the literature, this is referred to as a Real Amplitudes circuit ansatz, and is commonly used for quantum chemistry applications. This is because a rotation around the y-axis does not change the imaginary component of the complex amplitudes of a state vector, maintaining them as real, if they start as such.

\subsubsection{Variations}
\label{subsec:var}
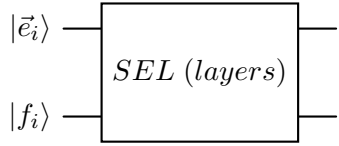
\begin{figure}[!htb]
    \centering
    \begin{quantikz}
        \lstick{$\ket{\vec{e}_i}$}&\gate[2]{SEL \;(layers)}&\\
        \lstick{$\ket{f_i}$}&&\\
    \end{quantikz}
    \caption{Quantum circuit of the strong collision ansatz}
    \label{fig-strong}
\end{figure}

\begin{figure}[!htb]
    \centering
    \begin{quantikz}
        \lstick{$\ket{\vec{e}_i}$}&\gate[1]{SEL \;(layers)}&\gate[2]{\text{$C_{\vec{e}_i,\forall i}U3_i$}}&\\
        \lstick{$\ket{f_i}$}&&&\\
    \end{quantikz}
    \caption{Quantum circuit of the strong-CU3 collision ansatz}
    \label{fig:strong-cu3}
\end{figure}
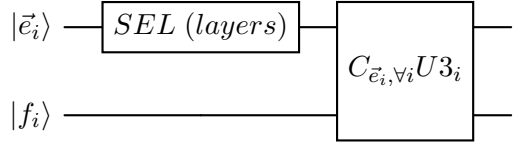

A description of each variation of the collision evaluated is introduced below. The naming of the ansatz has been adopted by the authors, and is not to be confused with any similar names for different ansatz that may have appeared in the literature. All variations have at most a single ancilla qubit, which is reset after measurement. The ancilla is considered as a part of either the lattice basis vector (lattice index) register, $\ket{\vec{e}_i}$, or the discrete density value (value) register, $\ket{f_i}$, as specified under each respective variation.
\paragraph{Strong Ansatz}
The Strong ansatz consists of a single strong entangling layer acting on the lattice index and value register concurrently, as shown in Fig.~\ref{fig-strong}. When included, the ancilla is considered as a part of the value register. It is considered as a benchmark to compare against conjectured architecture. In the absence of an ancilla, and with increased depth, a strong ansatz acting on $n$ qubits approaches a generic $n$-qubit unitary.
\paragraph{Strong-CU3}
The Strong-CU3 ansatz, shown in Fig.~\ref{fig:strong-cu3}, consists of a strong entangling layer acting on the lattice index register, followed by $U3$ gates acting on the value register, controlled on each state of the lattice index; e.g., in $D1Q3$, three $U3$ gates control on each of the three state vectors $\ket{i = 0}$, $\ket{i = 1}$ and $\ket{i = 2}$, respectively. We have hitherto used the notation $C_{\vec{e}_i,\forall i}U3_i$to refer to the operation $\Pi_{\forall\;j\;\in\;[0,Q[}C_{\ket{\vec{e}_i}=\ket{\vec{e}_j}}U3(\theta_j,\phi_j,\delta_j)$, and a similar notation for each operation applied, and controlled on every state of the lattice index register separately.

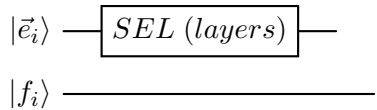
\begin{figure}[!htb]
    \centering
    \begin{quantikz}
        \lstick{$\ket{\vec{e}_i}$}&\gate[1]{SEL \;(layers)}&\\
        \lstick{$\ket{f_i}$}&&&\\
    \end{quantikz}
    \caption{Quantum circuit of the Scramble collision ansatz}
    \label{fig:Scramble}
\end{figure}

\paragraph{EL(2)-COP-Inverse-EL(2)}
The EL-COP-Inverse-EL ansatz consists of an EL and its inverse acting on the lattice index register, and sandwiching a $C_{\vec e_i,\forall i}OP$ operation on the value register. The EL2-COP-Inverse-EL variation includes an EL and its inverse also sandwiches the $C_{\vec e_i,\forall i}OP$ operation, acting on the value register. The EL could either be BEL or SEL. BEL2-CRY-Inverse-BEL2 is shown as an example in Fig.~\ref{fig:ucru}. Each of the ELs acting on the lattice index and value registers are of the same depth. When included, the ancilla is considered as part of the value register.
\begin{figure}[!htb]
    \centering
    \begin{quantikz}
        \lstick{$\ket{\vec{e}_i}$}&\gate[1]{BEL_{\text{$\vec{e}_i$}} \;(layers)}&\gate[2]{\text{$C_{\vec{e}_i,\forall i}RY_i$}}&\gate[1]{BEL\text{$^{-1}$}_{\text{$\vec{e}_i$}} \;(layers)}&    \\
        \lstick{$\ket{f_i}$}&\gate[1]{BEL_{\text{$f_i$}} \;(layers)}&&\gate[1]{BEL\text{$^{-1}$}_{\text{$f_i$}} \;(layers)}&\\
    \end{quantikz}
    \caption{Quantum circuit of the BEL2-CRY-Inverse-BEL2 ansatz, an example of the EL(2)-COP-Inverse-EL(2) family of ansatz}
    \label{fig:ucru}
\end{figure}
\paragraph{$2^{nd}$ Order EL(2)-COP-Inverse-EL(2) or EL(2)-COP-EL-COP-Inverse-EL(2)}
A second order variation of the EL(2)-COP-Inverse-EL(2) system involves setting up two identical circuits of the latter, with the same weights. The COP acting on each value register qubit is split into two (half angles are used twice), and the two sets of COPs are then separated by an EL acting on the two value registers of each copy of this circuit. It is constructed to explicitly allow second-order nonlinear operations on the values of the discrete densities. Higher-order variations could be constructed by providing more than two copies of the circuit. The $2^{nd}$ order variation of the BEL2-CRY-Inverse-BEL2 ansatz, presented in Fig.~\ref{fig:ucru}, is shown in Fig.~\ref{fig:el2nd}.
\begin{figure}[!htb]
    \centering
    \begin{quantikz}
        \lstick{$\ket{\vec{e}_i}$}&\gate[1]{BEL_{\text{$\vec{e}_i$}} \;(layers)}&\gate[2]{\text{$C_{\vec{e}_i,\forall i}RY_i$}}&&\gate[2]{\text{$C_{\vec{e}_i,\forall i}RY_i$}}&\gate[1]{BEL\text{$^{-1}$}_{\text{$\vec{e}_i$}} \;(layers)}&    \\
        \lstick{$\ket{f_i}$}&\gate[1]{BEL_{\text{$f_i$}} \;(layers)}&&\gate[2]{BEL(layers)}&&\gate[1]{BEL\text{$^{-1}$}_{\text{$f_i$}} \;(layers)}&\\
        \lstick{$\ket{f_i}$}&\gate[1]{BEL_{\text{$f_i$}} \;(layers)}&\gate[2]{\text{$C_{\vec{e}_i,\forall i}RY_i$}}&&\gate[2]{\text{$C_{\vec{e}_i,\forall i}RY_i$}}&\gate[1]{BEL\text{$^{-1}$}_{\text{$f_i$}} \;(layers)}&\\
        \lstick{$\ket{\vec{e}_i}$}&\gate[1]{BEL_{\text{$\vec{e}_i$}} \;(layers)}&&&&\gate[1]{BEL\text{$^{-1}$}_{\text{$\vec{e}_i$}} \;(layers)}&    \\
    \end{quantikz}
    \caption{Quantum circuit of the BEL2-CRY-Inverse-BEL2 ansatz, an example of the EL(2)-COP-Inverse-EL(2) family of ansatz}
    \label{fig:el2nd}
\end{figure}
\paragraph{Scrambler}
The scrambler ansatz is obtained by setting the $CU3$ rotation angles to be zero in the Strong-CU3 ansatz. As shown in Fig.~\ref{fig:Scramble}, a Scrambler ansatz has an SEL acting solely on the lattice index register. When included, the ancilla is considered part of the lattice index register.
\paragraph{Repetitions}
The final important notation for facilitating the description of the ansatz used is that of repetition. If the same circuitry is applied multiple times with different weights, we add a multiplier to the name of the circuit. For example, BEL-CRY-Inverse-BEL-BEL-CRY-Inverse-BEL refers to the application of the BEL-CRY-Inverse-BEL circuitry twice sequentially, with different weights each time. This would be written as (BEL-CRY-Inverse-BEL)$\times2$. In these instances, the number of layers describes the number of layers within one basic entangling layer (BEL) and not the total number of layers with the sequential application of the same circuitry multiple times.

\subsubsection{Summary of Key Parameters}
We present the key parameters of the major variations of the collision ansatz considered in Tab.~\ref{tab:sumparam}. The ansatz column follows the naming convention introduced in Sec.~\ref{subsec:var}. The $1-$ and $2-$Qubit gate counts provide the total gate count for the collision ansatz in terms of other parameters, the number of layers $L$, the lattice index register qubit count $qc$, and the number of repetitions $R$, where applicable. The circuit depth listed below is an upper bound. The actual circuit depth observed are lower by a factor $O(2-10)$ due to the varying distance between the target and control qubits in the application of the two-qubit gates in the entangling layer (see Sec.~\ref{subsec:el}). Apart from the circuit depth reported here being an upper bound, all other parameters are exact rather than an order of magnitude estimates or bounds.
\begin{table}
    \centering
    \begin{tabularx}{\textwidth}{XXXXXXXX}
         Ansatz& 1-Qubit Gate Count & 2-Qubit Gate Count& Circuit Depth ($<$)&\# of Parameters\\
         BEL2-CRY-Inverse-BEL2 & $2L(qc+1)$& $2L(qc+1)+2^{qc}qc$ & $2L(qc+1)+2^{qc}qc$&$L(qc+1)+2^{qc}$\\
         SEL2-CU3-Inverse-SEL2  & $2L(qc+1)$ & $2L(qc+1)+2^{qc}qc$ & $2L(qc+1)+2^{qc}qc$&$3(L(qc+1)+2^{qc})$\\
         (SEL2-CU3)$\times R$& $RL(qc+1)$ & $R(L(qc+1) +2^{qc}qc)$& $RL(qc+1)+R2^{qc}qc$&$3R(L(qc+1)+2^{qc})$ \\
 SEL-CRY-Inverse-SEL&$2Lqc$&$2Lqc+2^{qc}qc$&$2L(qc+1)+2^{qc}qc$&$3Lqc+2^{qc}$ \\
 $2^{nd}$ Order SEL-CRY-Inverse-SEL& $2Lqc+4b^2$ & $2Lqc+2^{qc+1}qc+2$ &$2L(qc+1)+2^{qc+1}qc+4$& $3Lqc+2^{qc+1}+12$  \\
 Strong& $L(qc+1)$ & $L(qc+1)$&$L(qc+2)$&$3L(qc+1)$  \\
    \end{tabularx}
        \caption[Summary of key parameters of the major variations of the collision ansatz considered]{Summary of key parameters of the major variations of the collision ansatz considered in this study, $L$, number of layers of the entangling layer as explained in Sec.~\ref{subsec:el}, and $qc$ number of qubits in the lattice index register $\ceil{\log_2{Q}}$}
    \label{tab:sumparam}
\end{table}
The Strong ansatz is the most favorable, being bilinear in all the parameters on which it depends. This is due to the lack of the rotations acting on the value register controlled on each index encoded in the index register. The latter adds an exponential term in $qc$. The consideration of the strong instead of basic entangling layer triples the number of parameters offered by the layer at no cost for the circuit depth. This is due to the $U3$ gate being native in the simulator. In practice, this gate is not part of the set native to a hardware platform, and a similar increase in the depth is expected for one-qubit gates as the $U3$ gate needs to be decomposed. Increasing the size of the index register increases the number of parameters the most, with an exponential scaling in the number of qubits $qc$, but it equally penalizes the circuit depth. We will refer again to these complexities in the context of the results presented in Sec.~\ref{sec:res}
\subsubsection{Hyperparameters}

\paragraph{Samples and Batch Sizes}
For exploratory purposes, the sample size has been set to $32$, while $100,000$ samples were used for training the final circuit. The batch size has been fixed to $32$ throughout. 
\subparagraph{Number of Epochs}
The number of epochs was set to $200$  initially with no early stopping criteria. Where promise is seen, additional epochs up to $1000$ total epochs have been used, as indicated.

\paragraph{Learning Rate}
The training has been started consistently with a learning rate of $0.3$ for which all experiments showed convergence. The learning rate has been reduced by one and two orders of magnitude consecutively in an attempt to fine-tune the trained circuit with no appreciable gain. 
\paragraph{Optimizer}
Stochastic gradient descent (SGD) and adaptive moment estimation (Adam) have been tested for a particular choice of optimizer, and all the results presented in this chapter are obtained using this choice.

\paragraph{Over-fitting a Small Dataset}
We start by attempting to overfit it with a small training set of only 32 samples of pre- and post-collision discrete densities. We use the random sample generation algorithm used in \cite{corbettaLearningLatticeBoltzmann2023}. 

\subsection{Lattice Boltzmann Symmetries}
To improve the performance of the neural network performing the collision, \cite{corbettaLearningLatticeBoltzmann2023} suggests hardwiring the collision invariants and equivariances, namely:
\begin{itemize}
    \item Scale equivariance
    \item Rotation and reflection equivariance
    \item Mass and momentum invariance
    \item Positivity of discrete densities
\end{itemize}
We shall consider each of them briefly below.

\subsubsection{Scale Equivariance}

Reference \cite{corbettaLearningLatticeBoltzmann2023} imposes scale equivariance by normalizing the discrete densities by the local fluid density before performing the collision. The resulting post-collision status of discrete densities is then denormalized by multiplying by the density again. On a quantum computer, performing this operation explicitly is challenging as it introduces further nonlinearity. The density, after all, is the sum of the discrete densities. We do not consider the scale equivariance explicitly in the construction of the quantum neural network.
\subsubsection{Rotation and reflection equivariance}
We are bound to the same D2Q9 lattice considered in \cite{corbettaLearningLatticeBoltzmann2023} and its corresponding symmetry group. The eighth-order dihedral symmetry group,
\begin{equation}
    D_8 = {\hat{I}, \hat{r}, \hat{r}^2, \hat{r}^3, \hat{s}, \hat{r}\hat{s}, \hat{r}^2\hat{s}, \hat{r}^3\hat{s}},
\end{equation}
is generated by a right-angle rotating transform $\hat{r}$, and a mirroring transform $\hat{s}$. In plain terms, it rotates the discrete densities in $90$ degree increments, as well as the reflection of the discrete densities about the x- and y-axes, as well as the first and second bisectors (the diagonals).  

\subsubsection{Mass and momentum invariance}
One may define mass and momentum as additional outputs of the neural network. Alternatively, as we have done already, error in mass and momentum amplitude may be incorporated into the loss function during the training of the quantum neural network.
\subsubsection{Positivity of discrete densities}
In our approach, strict positivity is enforced by construction of the amplitude encoding choice.
\subsection{Symmetry Preservation}
\label{sec:sympres}
To hard-wire the rotation and reflection equivariance into the neural network architecture, one may introduce additional qubits by which swap and rotation operations on the lattice basis vector register are controlled. For the purpose, the additional qubits are initialized into superposition, by application of Hadamard gates. The cardinality of the symmetry group for each second-order accurate lattice, as well as the additional qubit requirement, $\ceil{\log_2{C}}$, is detailed in Table~\ref{tab:cardsym}.
\begin{table}[!htbp]
        \caption[Cardinality and qubit count for the rotation and reflection equivariance symmetry group in different lattices]{Cardinality (C) and qubit count for the rotation and reflection equivariance symmetry group in each $D$ dimensional lattice}
\begin{tabularx}{\textwidth}{|X|X|X|}
  \hline
  Lattice & C & \# of Qubits
  \\\hline
  D1Q3 & 2 & 1
  \\\hline
  D2Q9 & 8 & 3
  \\\hline
  D3Q27 & 48 & 6
  \\\hline
  \end{tabularx}
  \label{tab:cardsym}
\end{table}
Then, one applies the $D_8$ transformations to the input discrete density vector, performing the collision, applying the inverse transformations, and computing each discrete density averaged over the $8$ resulting values. Let $\hat{\sigma} \in D_8$, and $\hat{N}$ be the action of the collision neural network. We then have
\begin{equation}
    \vec{\Tilde{f}}^{post} = \frac{1}{8}\Sigma_{\hat{\sigma}\in D_8} \hat{\sigma}^{-1}\hat{N}\hat{\sigma} \vec{f}^{pre},
\end{equation}
where the post and pre superscripts indicate the post-collision and pre-collision densities, respectively, and the $\vec{\Tilde{f}}^{post}$ is the vector of post-collision densities approximated by the neural network, as opposed to the exact $\vec{f}^{post}$ one. For the application of the symmetry transforms we relax the use of the encoding of the lattice basis vector described above, in favor of the traditional typical basis encoding $\ket{\vec e_i} \rightarrow \ket{i}$, e.g. $\ket{\vec e_3}\rightarrow\ket{3} = \ket{0011}$. This allows the definition of a unitary consisting solely of tensored and linearly combined Hubbard projectors.

The hard-wiring of the $D_8$ symmetries not only amount to an effective increase of the training dataset by $8$, but also improve the conservation of the pre-collision density, as well as velocity magnitudes in the post-collision predictions of the neural network.

\section{Results}
\label{sec:res}

\paragraph{Nonlinear Collision Operator}
 A collision ansatz using the modified amplitude encoding described above benefits from a reduction in the qubit complexity since only a single qubit is required for the value register. An SEL-CRY-Inverse-SEL collision ansatz has been used, as shown in Tab.~\ref{tab:amp-nonsym}. 

The cases considered with $16$ and $20$ layers with $192$ and $240$ parameters, respectively, achieve an order of magnitude improvement in the loss within $10$ epochs of training. We highlight again that with the amplitude encoding, it was practical to set a sample size of nearly $O(10^6)$, such that the processed count is $O(10^7)$ at $10$ epochs. While Tabs.~\ref{tab:amp-nonsym}~-~\ref{tab:amp-nonsym-2} start reporting at $10$ epochs for consistency, a loss value of $O(10^{-2})$ was indeed achieved within the first epoch of training with amplitude encoding. The final loss value of $O(10^{-3})$ was achieved with a processed count $O(10^8)$. 

Moreover, we also demonstrate the training results of a $2^{nd}$ Order SEL-CRY-Inverse-SEL ansatz in Tab.~\ref{tab:amp-nonsym-2}. The transform and inverse transform (using the same weights) are applied to each copy of the index register, whereas the controlled rotation acting on the value register is split and separated by an entangling layer acting on the latter. This amounts to only $24$ additional parameters in spite of doubling the qubit count. A similar loss of $O(10^{-3})$ was achieved with a training cost of $O(10^{7})$ processed count, with the case of $20$ layers stagnating at a loss of $O(10^{-1})$. As such, construction of higher-order ansatzes by including copies of the circuitry and allowing their value registers to interact is not investigated further.

\paragraph{Training with $D_8$ Symmetry Transforms}
The D8 symmetry transforms form a data augmentation technique. In \cite{corbettaLearningLatticeBoltzmann2023}, $8$ copies of the dataset are transformed, including one that is acted upon by the identity operator, which is also a member of the $D_8$ transforms. The corresponding inverse transform is then applied to each set of the predicted discrete density values. Finally, the predicated values are averaged. In this work, we leverage the parallelism allowed by the paradigm of quantum computing for a slightly different approach. The transforms are applied in-situ during training by hard-wiring the transforms and inverse transforms into the quantum circuit. This is done by including $\log_2 8 = 3$ ancilla qubits upon which we control the application of the unitary, which corresponds to one of the transforms, onto the lattice index register. It suffices here to say that the transforms amount to switching the discrete densities among themselves, which could be performed unitarily as it is reversible and entails no loss of information; the interested reader referred to available codes for details of this implementation. The averaging is handled by tracing over these ancillas, only by measuring the index and value registers. 

This might seem like a cosmetic improvement over directly augmenting the dataset. However, when a quantum simulator on a classical computer is considered, this amounts to augmenting the dataset without increasing the number of quantum circuit evaluations required. We recall here that the number $8$ is simply the cardinality of the relevant symmetry group for the $D2Q9$ lattice, which is $2$ and $48$ for $D1Q3$ and $D3Q27$, respectively.

Having settled on a collision ansatz, we also increase the dataset size to $8$ million ($8,388,608$). The number of layers is also increased to the maximum possible with our simulation setup, considering the availability of server nodes for training, and the time taken for the study. The number is chosen to be an even multiple of $2^4$ since $4$ qubits are used for the lattice index register, as to allow for a full exchange of information among the discrete densities.

The results of training the SEL-CRY-Inverse-SEL ansatz with $64$ layers and the $D8$ symmetry transforms hard-wired into the circuitry are shown in Tab.~\ref{tab:symamp}. We achieve a loss of $3.96 \times 10^{-4}$ with $200$ epochs of training, for a processed count of $O(10^8)$, or $25$ times the dataset size.

For comparison, we have also trained a Strong ansatz with amplitude encoding and symmetry transforms, as shown in Tab.~\ref{tab:symstrongamp}, which stagnated at a loss value of $0.00374$ at $100$ epochs, nearly double that achieved with $100$ epochs of training with the proposed SEL-CRY-Inverse-SEL architecture, and $10$ times the final loss to which the latter converges.

\subsubsection{Complexity Analysis}
We reemphasize here that ideally the training cost must be accounted once for each model and lattice combination considered, because of the portability of the collision model considered. As of now, the parametrization of the circuit in terms of the single relaxation time used is not considered, and the latter is explicitly fixed by the value chosen in the generation of the training dataset. The number of qubits used also differs between training and implementation because of the $\ceil{\log_2{C}}$ ancilla qubits needed to implement the data augmentation strategy during training. The qubit and gate complexities of a single collision with the model are then
\begin{equation}
    \ceil{D\log_2{3}}+1
\end{equation}
qubits, and
\begin{equation}
    (2^3(\ceil{D\log_2{3}}+2^{\ceil{D\log_2{3}}})+1)2^{\ceil{D\log_2{3}}}
\end{equation}
gates. The number of gates is a logarithmic improvement over that achieved in \cite{itaniQuantumAlgorithmLattice2024}. Moreover, this approach eliminates the need for ancilla qubits, which were needed in \cite{itaniQuantumAlgorithmLattice2024} to perform Hamiltonian simulation. As for the number of gates, the complexity obtained here , which is $O(Q^3)$, is $6$ orders of magnitude less than what we would expect of \cite{itaniQuantumAlgorithmLattice2024}, $O(Q^9)$. An additional advantage is our ability to use the exact streaming methodology that utilizes the adder gate. Several variations of the latter exist, the most basic one of which requires additional number of qubits logarithmic in the grid size as $O(\log_2{G}$), and a gate complexity $O(D\log_2^2{G})$. Thus, the overall algorithm complexity, accounting for neither the training cost nor the cost of initialization, becomes
\begin{equation}
    O(2D+\log_2{(G)}+1)
\end{equation}
qubits and
\begin{equation}
    T((2^3(\ceil{D\log_2{3}}+2^{\ceil{D\log_2{3}}})+1)2^{\ceil{D\log_2{3}}}+D\log_2^2{G}) = O(TQ^3+TD\log_2^2G)
\end{equation}
gates. The overall gate complexity of the algorithm utilizing a learnt collision operator differs from those reported \cite{liPotentialQuantumAdvantage2024} for an approach utilizing classical Carleman linearization followed by the quantum linear system algorthim (QLSA), in so far as it depends on $Q^3$ instead of $\log_2{Q^3}$. However, using a minimum bound estimate for the gate complexity of the latter, and for the same precision, the current approach breaks even with the gate complexity of \cite{liPotentialQuantumAdvantage2024} at $G = O(10^5)$ in $2D$, and $G = O(10^{22})$ in $3D$. The advantage of \cite{liPotentialQuantumAdvantage2024} is consistency. Note that generalizing the complexities for explicit dependence on $\frac{\Delta t}{\tau}$ and the error $\varepsilon$ requires further analysis. We provide the basis for it in Sec.~\ref{sec:erroranalysis}.

\subsubsection{Error Analysis}
\paragraph{Relation between the Reynolds Number and Number of Time steps Required}
The questions we are trying to answer are: how many time steps are needed before the QNN error dominates over the inherent error of the method? What flow regimes can be simulated within the time span provided? We start by expressing the Reynolds number, $Re$, in terms of the number of lattice sites along a side of a square domain $N$.

The number of lattice sites along a dimension $N$ typically scales as $N \propto O(Re^{\frac{3}{4}})$ for the standard model using the BGK collision \cite{houLatticeBoltzmannSubgrid1994}. We may also proceed to write the Reynolds number in terms of the Mach number:
\begin{equation}
    Re = \frac{UL}{\nu} = \frac{UN\Delta x}{c_s^2\Delta t (\frac{\tau}{\Delta t} -\frac{1}{2})} = \frac{MaN}{c_s(\frac{\tau}{\Delta t} -\frac{1}{2})} = \frac{Ma}{c_s} \frac{1}{\frac{\tau}{\Delta t} -\frac{1}{2}} N.
\end{equation}
The condition of compressibility requires $Ma << 1$, with weak compressibility usually taken as $Ma < 0.1$. This provides a stricter bound for $Re$ for a given $N$. For $N = 32,64, 128, 256, 512$ and $1024$, we obtain $Re = 6, 11, 22, 44, 89$ and $177$, respectively. Typical fluid flow simulations could be measured in flow through time (FTT) as $10-20$ FTT, with simpler steady-state problems taking $5-10$ FTT, and more complex problems with large instabilities $50-100$ FTT. For the smallest $N$ considered, we can expect the total simulation to be $O(10^2)$, and $O(10^4)$ for the largest. We proceed to estimate the Reynolds number our trained QNN could handle by developing an expression for the error accumulation in time.

\paragraph{Error Propagation}
\label{par:error}
Having estimated the total number of required time steps $T$, we proceed to analyze the propagation of the error. Since the QNN was trained on a dataset of single collision steps, we expect the errors to accumulate as the solution evolves. Though the streaming step is exact, it redistributes the errors within the local discrete densities to neighboring cells, as
\begin{equation}
    \sigma^2_{f_i}(t+1) = (1-\frac{\Delta t}{\tau})^2\sigma^2_{f_i}(t)+(\frac{\Delta t}{\tau} \partial_\rho f_i^{eq})^2\sigma_\rho^2(t)+2(\frac{\Delta t}{\tau} \partial_{\norm{u}} f_i^{eq})^2\sigma_{\norm{u}}^2(t)
\end{equation}
with
\begin{equation}
    \sigma_{\norm{u}}^2 (t+1) = Q\sigma_{f_i}^2+b_2,
\end{equation}
and
\begin{equation}
    \sigma_{\rho}^2 (t+1) = Q\sigma_{f_i}^2+b_1.
\end{equation}
Solving the recursion gives an expression for the error term of each discrete density as
\begin{equation}
    \sigma^2_{f_i}(t) = B\frac{1-A^{t}}{1-A} \forall t \ge 0,
\end{equation}
where
\begin{equation}
    A = (1-\frac{\Delta t}{\tau})^2+Q(\frac{\Delta t}{\tau})^2((\partial_\rho f_i^{eq})^2+2(\partial_{\norm{u}}f_i^{eq})^2)
\end{equation}
and
\begin{equation}
    B = (\frac{\Delta t}{\tau})^2((\partial_\rho f_i^{eq})^2 b_1^2+2(\partial_{\norm{u}}f_i^{eq})^2b_2^2).
\end{equation}
The partial derivatives appearing could be expressed in terms of the Mach number as
\begin{equation}
    \partial_\rho f_i^{eq} = w_i (1+2Ma-\frac{1}{2}Ma^2),
\end{equation}
and
\begin{equation}
    \partial_{\vec u} f_i^{eq} = w_i \frac{\rho}{c_s}(1-Ma).
\end{equation}
To study the limiting case of $\sigma_{f_i}^2$, we observe that
\begin{equation}
    \partial_A  \sigma^2_{f_i} = B(A-1)^{-2}(tA^{t-1}(A-1)-A^t) \begin{cases}
        < 0 & t \le 1
        \\> 0 & t > 1
    \end{cases}
\end{equation}
, meaning the error term increases with an increasing A for all times starting the first timestep, since $1<\frac{A}{A-1}<2 \;\forall\;0<Ma<0.1$. We may then use the value of $A(Ma = 0)$ as an upper bound without loss of generality to conclude that the error from our method of approximating the collision dominates over the $O(Ma^2)$ error of the lattice Boltzmann method after $T_{max} \approx 230$ time steps, consistent with $Re \approx 6$.

The analysis above assumes that the error in distinct discrete densities is largely correlated through the hydrodynamic variables, $\rho$ and $\vec u$. The other end of the spectrum is random and concentrated in a single, unspecified discrete density. As a bounding case, we may consider the RMSE obtained above, $(3.69 \times 10^{-4})$, as an additive error for a single discrete density. The square of this term would then be added to $B$ in the formulation above, resulting in the solver diverging in $O(1)$ steps. 

We have trained our QNN with $\frac{\Delta t}{\tau} =1$. However, one could see from the propagation of error that setting $\frac{\Delta t}{\tau} < 1$ would directly improve the error divergence. We note that both $A$ and $B$ show dependence on $\frac{\Delta t}{\tau}$, and that $A<1$ for $2\times10^{-16}\le \frac{\Delta t}{\tau} \le 1.375726 \times 10^{-1}$. Or, practically,
\begin{equation}
    A < O(10^{-1}).
\end{equation}
This reduction corresponds to an increase in the relaxation time by roughly $10$, which corresponds to a similar decrease in the corresponding Reynolds number. This means that the $~230$ time steps now corresponds to $Re \approx 0.6$, and $Re \approx 6$ now requires $~2,300$ time steps. This would increase the range of applicability to $Re = O(10)$. Achieving an RSME of $O(10^{-5})$ would theoretically extend the application's range of applicability to $T_{max} = O(10^4)$, or $Re = O(10^3)$, and an RSME of $O(10^{-6})$ to $T_{max} = O(10^6)$ or $Re = O(10^5)$.

\begin{figure}
    \centering
    \includegraphics[width=0.8\columnwidth]{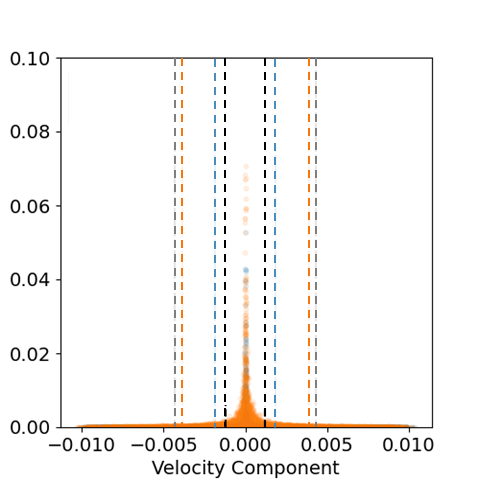}
    \caption[Error ratio of the velocity components post-collision]{The post-collision error ratio of the x-(blue) and y-(orange) velocity components is shown against the value of the respective component. The vertical dashed lines in blue, orange and grey represent the x-, y-component and combined standard deviation of the error distribution.}
    \label{fig:velerror}
\end{figure}

We note that our QNN effectively provides a unitary operator to approximate $\vec{f}^{eq}$ (for a given velocity range) which could be used in conjunction with the identity operator to construct a circuit for an arbitrary $\frac{\Delta t}{\tau}$ using the well-established quantum singular value transformation (QSVT) or linear combination of unitaries (LCU).
\label{sec:erroranalysis}
\paragraph{Lower Bound for Applicable Velocity Range}
\label{sec:lowbound}
While Sec.~\ref{sec:erroranalysis} presented a generalized error analysis, further inspection reveals that the error introduced into the velocities is dependent on its value. As expected, and shown in Fig.~\ref{fig:velerror}, the error ratio shoots over the true velocity value as the velocity approaches zero.
Figure~\ref{fig:velerror} shows the post-collision error ratio of the x-(blue) and y-(orange) velocity components against the value of the respective component. The vertical dashed lines in blue, orange and grey represent the x-, y-component and combined standard deviation of the error distribution. Their respective absolute values are $1.9\times 10^{-3}$, $3.9\times 10^{-3}$ and $4.3\times 10^{-3}$, respectively. The dashes black vertical line represents the unscaled training loss, $3\times RSME = 3\times 3.964\times10^{-4}$. 

It is intuitive in retrospect that the error is exaggerated for the smaller velocity values. To this end, we recall that when solving for fluid flow using lattice methods, the velocity is restricted to be smaller than unity. Therefore, smaller velocities benefit from the separation of scales introduced by the polynomial nonlinearity, e.g. $10^{-5} \gg (10^{-5})^2 = 10^{-10}$. As such, we use the combined standard deviation of the error ratio distribution, $4.3\times 10^{-3}$ as a lower cutoff for the applicable velocity range of our collision model.

\subsection{Demonstration of the Lid-Driven Cavity Flow}
To demonstrate the utility of our treatment of the collision operator, we should choose a suitable benchmark. We choose the lid-driven cavity problem for this purpose. The cavity flow case is commonly used for benchmarking numerical solvers for fluid dynamics for several reasons. Despite its simple geometry, the boundary conditions of the moving lid as well as the wall conditions on the sides create challenging conditions --- particularly around the corners --- for the stability of numerical solvers. Moreover, the case does not admit an analytical solution while having intricate flow features even at relatively low Reynolds number. The primary vortex is the main characteristic of the flow configuration, while the secondary and tertiary vortices develop around the corners for $Re = O(100)$. This has led to a wide adoption of the lid-driven cavity as a benchmark case and resulted in wide availability of datasets for comparison.

\subsubsection{Limitations of the Demonstration}
The overall workflow is based on conditionally inserting the quantum collision operator into an otherwise classical workflow. We have already established in Sec.~\ref{par:error} that the solver diverges in a handful of steps. We have seen in Sec.~\ref{sec:lowbound} that there are two mechanisms from which the error arises, one related to approximating the nonlinearity inherent in the collision which we expect to worsen with higher Reynolds number, and another is due to the lack of a computational zero in the amplitude encoding, as opposed to the binary encoding. To be able to evaluate how well the nonlinearity could be captured by the learnt term, we resort to using the cutoff introduced in Sec.~\ref{sec:lowbound}, artificially suppressing the error, to yield a convergent solution to assess.

\paragraph{Collision}
The collision is performed selectively with the quantum and classical collision operators based on whether the velocity magnitude at the site is larger or smaller than the cutoff calculated in Sec.~\ref{sec:lowbound}. It is this cutoff that renders our demonstration hybrid in its actuality. That is, the discrete densities are measured at every step, the streaming and boundary conditions are applied classically, and the post-collision densities are reinitialized.

Thus, the method introduced in this work is better adapted for constructing a nonlinear ``corrector" of a linear collision model as has been done for turbulence subgrid modeling \cite{ortaliKineticDatadrivenApproach2024} . This requires further work that combines the quantum algorithm considered here with one of those that presents a quantum algorithm for a lattice Boltzmann model with linear collision, of which we commend \cite{wawrzyniakUnitaryQuantumAlgorithm2024}. 

\paragraph{Streaming}
The discrete densities are measured at every timestep and streamed classically. Since our quantum collision operator is compatible with the existing quantum implementation of the exact streaming, and the encoding we use does not require renormalization of our variables, this process is valid without any further implications in regard to our claim to a self-contained quantum algorithm. This statement is not general and is valid only for the selected, SEL-CRY-Inverse-SEL, ansatz. 

This workflow amounts to problem-specific quantum circuit cutting. This process allows us to transmute what would ordinarily be a vast quantum calculation into a harmonious collection of smaller circuits, each amenable to execution with a simulation of smaller qubit count. The virtue of this technique lies in its capacity to bisect quantum circuits at strategically chosen junctures, introducing a classical intermediary between the quantum simulations. This necessitates the measurement of certain quantum states and their subsequent reinitialization.
\paragraph{Boundary Conditions}
Likewise, the boundary conditions are implemented classically. The boundary conditions at the wall amount to an exchange of discrete densities that could otherwise be performed unitarily in a complete quantum algorithm. Indeed, the implementation of this boundary condition is the same as the implementation of the D8 symmetries that has been already done. The constant velocity at the lid is obtained by relaxing the discrete densities to known constant values precalculated on the basis of the fixed lid velocity. As such, the classical implementation of the boundary conditions is not a fundamental limitation of the demonstration either.
 
\subsubsection{Key Parameters}
The Mach number is set to $Ma = 0.1$, whereas the Reynolds number is determined by trial and error from the convergence of the quantum collision case with the cut off as $Re = 40$. The cases for $Re = 10$ and $Re = 40$ are presented here, each with an appropriately chosen grid size. The reference length of the domain is kept at unity.

\subsubsection{Results and Discussion}
\begin{figure}
     \centering
     \begin{subfigure}[b]{0.5\textwidth}
         \centering
         \includegraphics[width=\textwidth]{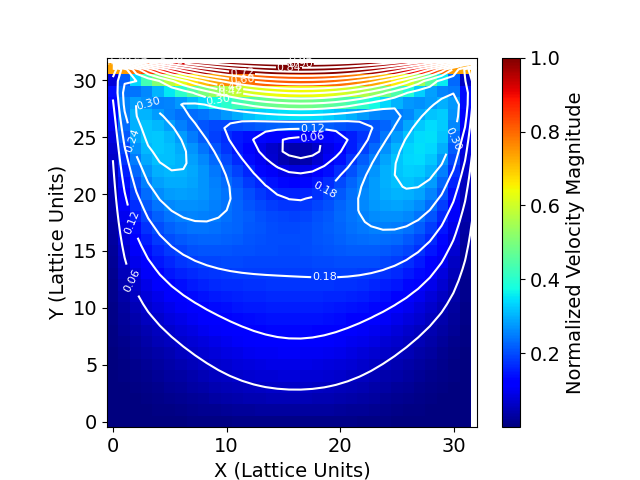}
         \caption{Classical}
         \label{fig:10clb}
     \end{subfigure}
     \\
     \begin{subfigure}[b]{0.5\textwidth}
         \centering
         \includegraphics[width=\textwidth]{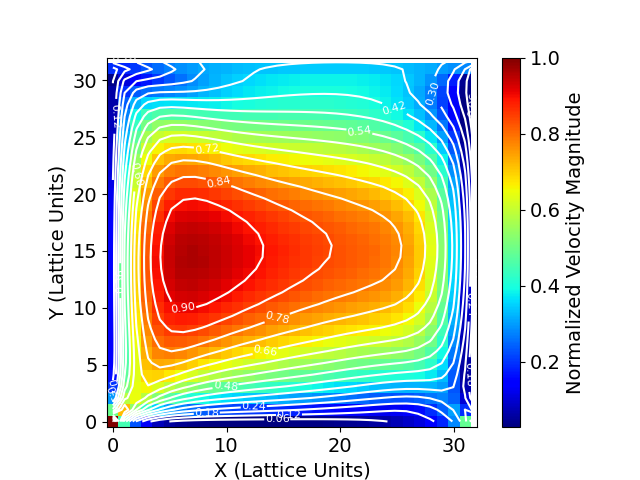}
         \caption{Without cutoff}
         \label{fig:10unc}
     \end{subfigure}
     \\
     \begin{subfigure}[b]{0.5\textwidth}
         \centering
         \includegraphics[width=\textwidth]{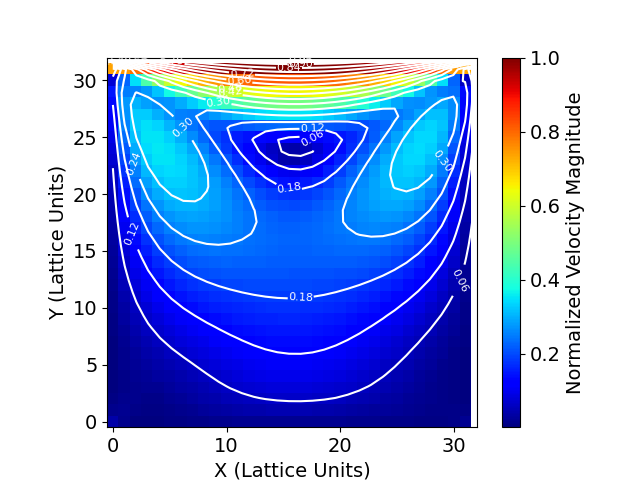}
         \caption{With cutoff}
         \label{fig:10qlb}
     \end{subfigure}
        \caption[Solution of the lid-driven cavity at $Re = 10$ on a $32\times 32$ grid]{Normalized velocity magnitude corresponding to the flow field of the lid-driven cavity at Re = 10 solved using the lattice Boltzmann method with \ref{fig:10clb} classical, \ref{fig:10unc} quantum, and \ref{fig:10qlb} quantum collision applied only to lattice sites with velocity magnitudes larger than the cutoff. The white contours represent the isovelocity lines.}
        \label{fig:Re10}
\end{figure}

\begin{figure}
     \centering
     \begin{subfigure}[b]{0.5\textwidth}
         \centering
         \includegraphics[width=\textwidth]{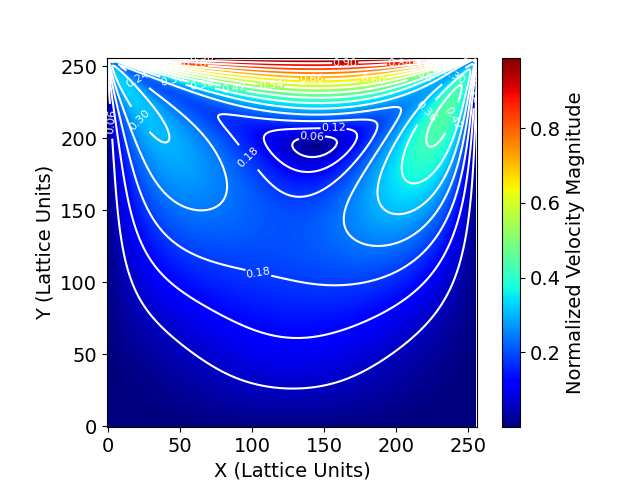}
         \caption{Classical}
         \label{fig:40clb}
     \end{subfigure}
     \\
     \begin{subfigure}[b]{0.5\textwidth}
         \centering
         \includegraphics[width=\textwidth]{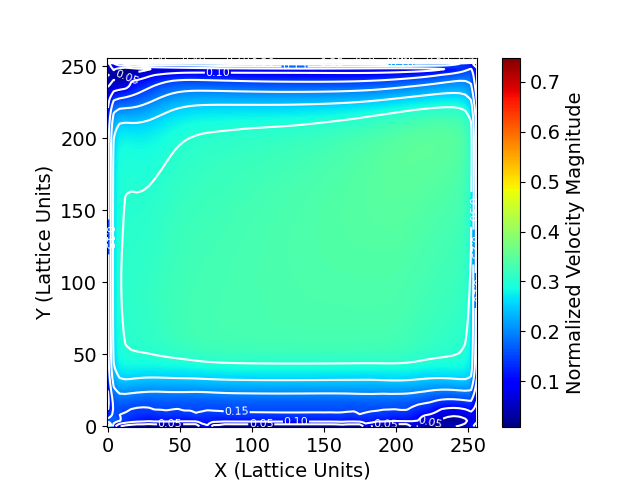}
         \caption{Without cutoff}
         \label{fig:40unc}
     \end{subfigure}
     \\
     \begin{subfigure}[b]{0.5\textwidth}
         \centering
         \includegraphics[width=\textwidth]{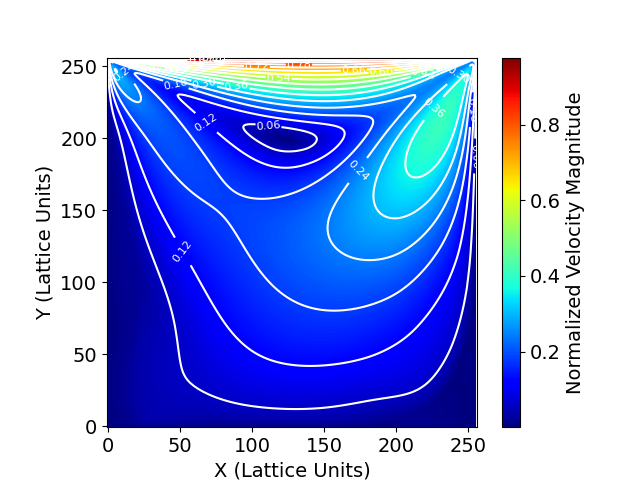}
         \caption{With cutoff}
         \label{fig:40qlb}
     \end{subfigure}
        \caption[Solution of the lid-driven cavity on a $256\times 256$ grid at $Re = 40$]{Normalized velocity magnitude corresponding to the flow field of the lid-driven cavity at Re = 40 on a $256\times 256$ grid solved using the lattice Boltzmann method with \ref{fig:40clb} classical, \ref{fig:40unc} quantum, and \ref{fig:40qlb} quantum collision applied only to lattice sites with velocity magnitudes larger than the cutoff. The white contours represent the isovelocity lines.}
    \label{fig:Re40}
\end{figure}

We show the solutions of the lid-driven cavity for $Re = 10$ and $Re = 40$, Figs.~\ref{fig:Re10}~and~\ref{fig:Re40}, respectively. The figures show the solutions using the quantum collision operator without and with the velocity cutoff introduced above. In both solutions without the cutoff, the flow field is subject to a diverging error emanating from the domain walls. This is in line with the visualization of the error in Fig.~\ref{fig:velerror} which shows a singular peak in the vicinity of null velocity. Upon initial inspection, it is evident that the cutoff quantum solution, at best, represents a distortion of the classical solution. With the cutoff quantum case, we observe a tertiary flow feature seemingly originating from the bottom left corner, permeating through the domain. This appears to be pressing upon the primary flow structure, in which we see the isovelocity contours pressed inwards compared to the classical case. This observation is true for both Reynolds numbers, in each of which the isovelocity contours of the main vortex are distorted in a similar fashion, though to different degrees. This is seen in the flattening of isovelocity contours of the main vortex in the bottom left corner. However, the effect plays out differently in the top left corner, with the high velocity streak enlarged in the case of $Re = 10$ but reduced for $Re = 40$.

These results are in line with the numerical instability arising from ill-conditioned domain and image of the nonlinear driving function which we truncated to achieve the higher Reynolds number. In other words, approximating the nonlinear function introduces an instability at the smaller rather than larger velocity values, and the error thereby plays the role of forcing at the small scale.

\section{Conclusion}
We have presented an attempt at learning a unitary operator to approximate the lattice Boltzmann collision operator. The hard-wiring of the lattice Boltzmann symmetries into the quantum circuit has been described. Contrary to the prevailing explanation, we have shown that approximating the nonlinear system is limited from below at velocity values comparable to the error of the method. We have demonstrated the collision step can be modified with amplitude encoding for velocities in the range of $O(10^{-2}-10^{-1})$ velocity lattice units. Finally, the quantum simulation of classical flows has been introduced as a surmountable problem for quantum machine learning, to allow for it to be tackled in a further cross-disciplinary fashion. 

Trade-offs have materialized differently for the streaming and collision steps. The parallelization of the streaming step has come from reconsidering the basis for the lattice index encoding, with an increase in the required qubit count, $1$ qubit in both the $2D$ and $3D$ cases. Moreover, we have shown that consideration of the symmetries of the lattice in a physically-motivated ansatz have each reduced the training loss by an order of magnitude. This balancing of the efficiencies and accuracies of the methods considered would not have been possible with the consideration of a generic solver, without turning attention to the numerics and the accuracy of the specific application.

Finally, a few words are in order on the application of the cavity flow chosen here to test our ideas in this paper. It is known to be a difficult flow for any simulation method, but we chose it to push our ideas as much as possible. We have learned that while the quantum machine learning is very efficient, its application is confined to a range of velocities bounded from below. The collision ansatz based on the modified amplitude encoding used in the demonstration is ill-conditioned to handle a wider range of velocities; thus, it is unlikely to benefit from simple interventions (e.g., increased circuit depth, etc). The encoding choice must be revisited, e.g. parameterized with an autoencoder approach. The QML approach shows promise, but still suffers from a training loss two orders of magnitude above what is necessary to achieve a quantum advantage over classical computers in the range of Reynolds numbers considered.

\acknowledgements
We would like to acknowledge  A. Ameri (MIT), S. S. Bharadwaj (NYU), L. Domingo (Ingenii), A. Gabbana (LANL), M. E. A. Mondal (UoRoch), A. Rodhiya (NYU), M. Shoaib (NSU),  R. Steijl (UoGlas), and S. Succi (IIT), J. Zylberman (CERFACS) for their illuminating discussions, as well as  R.~Baltaji (UIUC) for her input on the quantum neural network training setup and D.~Ren (UCLA) for his collaborative efforts that build on this work.

W. Itani performed part of this work during visits to the Institute of Pure and Applied Mathematics (IPAM) at the University of California, Los Angeles (UCLA) and the Plasma Science and Fusion Center (PSFC) at the Massachusetts Institute of Technology (MIT). The work was supported partly by the US National Science Foundation Award 24-600, Number 2534977, under the title, TRAILBLAZER Quantum Computing and Machine Learning for Fluid Dynamics Research.

We dedicate this article to the memory of Prof. Nuno Loureiro.

\section*{Data Availability}
The code developed for this study is available at \href{https://github.com/waelitani/qml-lbm-collision}{https://github.com/waelitani/qml-lbm-collision}.

\bibliographystyle{plainnat}
\bibliography{references}

@misc{schuldInnovatingMachineLearning,
	address = {University of California, Los Angeles},
	title = {Innovating machine learning with near-term quantum computing},
	language = {en},
	author = {Schuld, Maria},
	year = {2019}
}

@misc{billLatticeBoltzmannMethod,
	address = {Simon Fraser University},
	title = {Lattice {Boltzmann} {Method} for {Fluid} {Simulations}},
	language = {en},
	author = {Bill, Yuanxun and Meskas, Justin},
	year = {2011},
}

@article{steijlParallelEvaluationQuantum2018a,
	title = {Parallel evaluation of quantum algorithms for computational fluid dynamics},
	volume = {173},
	issn = {00457930},
	url = {https://linkinghub.elsevier.com/retrieve/pii/S0045793018301841},
	doi = {10/gd9qx3},
	language = {en},
	urldate = {2020-11-17},
	journal = {Computers \& Fluids},
	author = {Steijl, René and Barakos, George N.},
	month = sep,
	year = {2018},
	pages = {22--28},
}

@inproceedings{randlesPerformanceAnalysisLattice2013,
	address = {Cambridge, MA, USA},
	title = {Performance {Analysis} of the {Lattice} {Boltzmann} {Model} {Beyond} {Navier}-{Stokes}},
	isbn = {978-1-4673-6066-1 978-0-7695-4971-2},
	url = {http://ieeexplore.ieee.org/document/6569885/},
	doi = {10.1109/IPDPS.2013.109},
	language = {en},
	urldate = {2021-08-06},
	booktitle = {2013 {IEEE} 27th {International} {Symposium} on {Parallel} and {Distributed} {Processing}},
	publisher = {IEEE},
	author = {Randles, Amanda Peters and Kale, Vivek and Hammond, Jeff and Gropp, William and Kaxiras, Efthimios},
	month = may,
	year = {2013},
	pages = {1063--1074},
}

@article{itaniAnalysisCarlemanLinearization2022,
	title = {Analysis of {Carleman} {Linearization} of {Lattice} {Boltzmann}},
	volume = {7},
	copyright = {http://creativecommons.org/licenses/by/3.0/},
	url = {https://www.mdpi.com/2311-5521/7/1/24},
	doi = {10.3390/fluids7010024},
	language = {english},
	number = {1},
	urldate = {2022-01-12},
	journal = {Fluids},
	author = {Itani, Wael and Succi, Sauro},
	month = jan,
	year = {2022},
	note = {Number: 1
Publisher: Multidisciplinary Digital Publishing Institute},
	pages = {24},
}

@article{todorovaQuantumAlgorithmCollisionless2020a,
	title = {Quantum algorithm for the collisionless {Boltzmann} equation},
	volume = {409},
	issn = {0021-9991},
	url = {https://www.sciencedirect.com/science/article/pii/S0021999120301212},
	doi = {10.1016/j.jcp.2020.109347},
	language = {english},
	urldate = {2022-10-28},
	journal = {Journal of Computational Physics},
	author = {Todorova, Blaga N. and Steijl, René},
	month = may,
	year = {2020},
	pages = {109347},
}

@book{steijlQuantumAlgorithmsNonlinear2020,
	title = {Quantum {Algorithms} for {Nonlinear} {Equations} in {Fluid} {Mechanics}},
	isbn = {978-1-83968-134-9},
	url = {https://www.intechopen.com/state.item.id},
	language = {english},
	urldate = {2022-10-30},
	publisher = {IntechOpen},
	author = {Steijl, René},
	month = dec,
	year = {2020},
	doi = {10.5772/intechopen.95023},
	note = {Publication Title: Quantum Computing and Communications}
}

@article{corbettaLearningLatticeBoltzmann2023,
	title = {Toward learning {Lattice} {Boltzmann} collision operators},
	volume = {46},
	issn = {1292-895X},
	url = {https://doi.org/10.1140/epje/s10189-023-00267-w},
	doi = {10.1140/epje/s10189-023-00267-w},
	language = {en},
	number = {3},
	urldate = {2023-09-21},
	journal = {The European Physical Journal E},
	author = {Corbetta, Alessandro and Gabbana, Alessandro and Gyrya, Vitaliy and Livescu, Daniel and Prins, Joost and Toschi, Federico},
	month = mar,
	year = {2023},
	note = {Number: 3},
	pages = {10}
}

@article{pfefferReducedorderModelingTwodimensional2023,
	title = {Reduced-order modeling of two-dimensional turbulent {Rayleigh}-{Bénard} flow by hybrid quantum-classical reservoir computing},
	volume = {5},
	url = {https://link.aps.org/doi/10.1103/PhysRevResearch.5.043242},
	doi = {10.1103/PhysRevResearch.5.043242},
	number = {4},
	urldate = {2024-01-28},
	journal = {Physical Review Research},
	author = {Pfeffer, Philipp and Heyder, Florian and Schumacher, Jörg},
	month = dec,
	year = {2023},
	note = {Number: 4
Publisher: American Physical Society},
	pages = {043242}
}

@article{au-yeungQuantumAlgorithmSmoothed2024,
	title = {Quantum algorithm for smoothed particle hydrodynamics},
	volume = {294},
	issn = {0010-4655},
	url = {https://www.sciencedirect.com/science/article/pii/S0010465523002540},
	doi = {10.1016/j.cpc.2023.108909},
	urldate = {2024-01-28},
	journal = {Computer Physics Communications},
	author = {Au-Yeung, R. and Williams, A. J. and Kendon, V. M. and Lind, S. J.},
	month = jan,
	year = {2024},
	pages = {108909}
}

@article{gaitanCircuitImplementationOracles2024,
	title = {Circuit implementation of oracles used in a quantum algorithm for solving nonlinear partial differential equations},
	volume = {109},
	url = {https://link.aps.org/doi/10.1103/PhysRevA.109.032604},
	doi = {10.1103/PhysRevA.109.032604},
	number = {3},
	urldate = {2024-03-12},
	journal = {Physical Review A},
	author = {Gaitan, Frank},
	month = mar,
	year = {2024},
	note = {Number: 3
Publisher: American Physical Society},
	pages = {032604},
}

@misc{wawrzyniakUnitaryQuantumAlgorithm2024,
	title = {Unitary {Quantum} {Algorithm} for the {Lattice}-{Boltzmann} {Method}},
	url = {http://arxiv.org/abs/2405.13391},
	language = {en},
	urldate = {2024-06-12},
	publisher = {arXiv},
	author = {Wawrzyniak, David and Winter, Josef and Schmidt, Steffen and Indinger, Thomas and Schramm, Uwe and Janßen, Christian and Adams, Nikolaus A.},
	month = jun,
	year = {2024},
	note = {Issue: arXiv:2405.13391
arXiv:2405.13391 [quant-ph]},
}

@article{wangGPUImplementedLatticeBoltzmann2024,
	title = {A {GPU}-{Implemented} {Lattice} {Boltzmann} {Model} for {Large} {Eddy} {Simulation} of {Turbulent} {Flows} in and around {Forest} {Shelterbelts}},
	volume = {15},
	issn = {2073-4433},
	url = {https://www.mdpi.com/2073-4433/15/6/735},
	doi = {10.3390/atmos15060735},
	language = {english},
	number = {6},
	urldate = {2024-06-30},
	journal = {Atmosphere},
	author = {Wang, Yansen and Zeng, Xiping and Decker, Jonathan and Dawson, Leelinda},
	month = jun,
	year = {2024},
	note = {Number: 6
Publisher: Multidisciplinary Digital Publishing Institute},
	pages = {735}
}

@article{budinskiQuantumAlgorithmAdvection2021,
	title = {Quantum algorithm for the advection–diffusion equation simulated with the lattice {Boltzmann} method},
	volume = {20},
	issn = {1570-0755, 1573-1332},
	url = {http://link.springer.com/10.1007/s11128-021-02996-3},
	doi = {10/ghz65x},
	language = {english},
	number = {2},
	urldate = {2021-02-09},
	journal = {Quantum Information Processing},
	author = {Budinski, Ljubomir},
	month = feb,
	year = {2021},
	note = {Number: 2},
	pages = {57}
}

@article{yepezLatticeGasQuantumComputation1998,
	title = {Lattice-{Gas} {Quantum} {Computation}},
	volume = {09},
	issn = {0129-1831},
	url = {https://www.worldscientific.com/doi/abs/10.1142/S0129183198001436},
	doi = {10/b4hdtz},
	number = {08},
	urldate = {2021-02-12},
	journal = {International Journal of Modern Physics C},
	author = {Yepez, Jeffrey},
	month = dec,
	year = {1998},
	note = {Number: 08
Publisher: World Scientific Publishing Co.},
	pages = {1587--1596},
}

@article{steijlQuantumCircuitImplementation2023,
	title = {Quantum {Circuit} {Implementation} of {Multi}-{Dimensional} {Non}-{Linear} {Lattice} {Models}},
	volume = {13},
	copyright = {http://creativecommons.org/licenses/by/3.0/},
	issn = {2076-3417},
	url = {https://www.mdpi.com/2076-3417/13/1/529},
	doi = {10.3390/app13010529},
	language = {english},
	number = {1},
	urldate = {2023-01-12},
	journal = {Applied Sciences},
	author = {Steijl, René},
	month = jan,
	year = {2023},
	note = {Number: 1
Publisher: Multidisciplinary Digital Publishing Institute},
	pages = {529}
}

@article{schalkersImportanceDataEncoding2024,
	title = {On the importance of data encoding in quantum {Boltzmann} methods},
	volume = {23},
	issn = {1573-1332},
	url = {https://doi.org/10.1007/s11128-023-04216-6},
	doi = {10.1007/s11128-023-04216-6},
	language = {en},
	number = {1},
	urldate = {2024-11-30},
	journal = {Quantum Information Processing},
	author = {Schalkers, Merel A. and Möller, Matthias},
	month = jan,
	year = {2024},
	pages = {20}
}

@article{schalkersEfficientFailsafeQuantum2024,
	title = {Efficient and fail-safe quantum algorithm for the transport equation},
	volume = {502},
	issn = {0021-9991},
	url = {https://doi.org/10.1016/j.jcp.2024.112816},
	doi = {10.1016/j.jcp.2024.112816},
	number = {C},
	urldate = {2024-11-30},
	journal = {J. Comput. Phys.},
	author = {Schalkers, Merel A. and Möller, Matthias},
	month = jun,
	year = {2024}
}

@article{schuldEffectDataEncoding2021,
	title = {Effect of data encoding on the expressive power of variational quantum-machine-learning models},
	volume = {103},
	url = {https://link.aps.org/doi/10.1103/PhysRevA.103.032430},
	doi = {10.1103/PhysRevA.103.032430},
	number = {3},
	urldate = {2024-11-30},
	journal = {Physical Review A},
	author = {Schuld, Maria and Sweke, Ryan and Meyer, Johannes Jakob},
	month = mar,
	year = {2021},
	note = {Publisher: American Physical Society},
	pages = {032430}
}

@article{liuEfficientQuantumAlgorithm2021,
	title = {Efficient quantum algorithm for dissipative nonlinear differential equations},
	volume = {118},
	url = {https://www.pnas.org/doi/10.1073/pnas.2026805118},
	doi = {10.1073/pnas.2026805118},
	number = {35},
	urldate = {2024-11-30},
	journal = {Proceedings of the National Academy of Sciences},
	author = {Liu, Jin-Peng and Kolden, Herman Oie and Krovi, Hari K. and Loureiro, Nuno F. and Trivisa, Konstantina and Childs, Andrew M.},
	month = aug,
	year = {2021},
	note = {Publisher: Proceedings of the National Academy of Sciences},
	pages = {e2026805118}
}

@article{schuldIntroductionQuantumMachine2015,
	title = {An introduction to quantum machine learning},
	volume = {56},
	issn = {0010-7514},
	url = {https://doi.org/10.1080/00107514.2014.964942},
	doi = {10.1080/00107514.2014.964942},
	number = {2},
	urldate = {2024-11-30},
	journal = {Contemporary Physics},
	author = {Schuld, Maria and Sinayskiy, Ilya and Petruccione, Francesco},
	month = apr,
	year = {2015},
	note = {Publisher: Taylor \& Francis
\_eprint: https://doi.org/10.1080/00107514.2014.964942},
	pages = {172--185}
}

@article{ameriQuantumAlgorithmLinear2023,
	title = {Quantum algorithm for the linear {Vlasov} equation with collisions},
	volume = {107},
	url = {https://link.aps.org/doi/10.1103/PhysRevA.107.062412},
	doi = {10.1103/PhysRevA.107.062412},
	number = {6},
	urldate = {2024-11-30},
	journal = {Physical Review A},
	author = {Ameri, Abtin and Ye, Erika and Cappellaro, Paola and Krovi, Hari and Loureiro, Nuno F.},
	month = jun,
	year = {2023},
	note = {Publisher: American Physical Society},
	pages = {062412}
}

@article{itaniQuantumAlgorithmLattice2024,
	title = {Quantum algorithm for lattice {Boltzmann} ({QALB}) simulation of incompressible fluids with a nonlinear collision term},
	volume = {36},
	issn = {1070-6631},
	url = {https://doi.org/10.1063/5.0176569},
	doi = {10.1063/5.0176569},
	number = {1},
	urldate = {2024-11-30},
	journal = {Physics of Fluids},
	author = {Itani, Wael and Sreenivasan, Katepalli R. and Succi, Sauro},
	month = jan,
	year = {2024},
	pages = {017112}
}

@article{mezzacapoQuantumSimulatorTransport2015,
	title = {Quantum {Simulator} for {Transport} {Phenomena} in {Fluid} {Flows}},
	volume = {5},
	copyright = {2015 The Author(s)},
	issn = {2045-2322},
	url = {https://www.nature.com/articles/srep13153},
	doi = {10.1038/srep13153},
	language = {en},
	number = {1},
	urldate = {2024-11-30},
	journal = {Scientific Reports},
	author = {Mezzacapo, A. and Sanz, M. and Lamata, L. and Egusquiza, I. L. and Succi, S. and Solano, E.},
	month = aug,
	year = {2015},
	note = {Publisher: Nature Publishing Group},
	pages = {13153}
}

@article{kocherlaFullyQuantumAlgorithm2024,
	title = {Fully quantum algorithm for mesoscale fluid simulations with application to partial differential equations},
	volume = {6},
	issn = {2639-0213},
	url = {https://doi.org/10.1116/5.0217675},
	doi = {10.1116/5.0217675},
	number = {3},
	urldate = {2024-11-30},
	journal = {AVS Quantum Science},
	author = {Kocherla, Sriharsha and Song, Zhixin and Chrit, Fatima Ezahra and Gard, Bryan and Dumitrescu, Eugene F. and Alexeev, Alexander and Bryngelson, Spencer H.},
	month = sep,
	year = {2024},
	pages = {033806},
}

@article{giliIntroducingNonlinearActivations2023,
	title = {Introducing nonlinear activations into quantum generative models},
	volume = {107},
	url = {https://link.aps.org/doi/10.1103/PhysRevA.107.012406},
	doi = {10.1103/PhysRevA.107.012406},
	number = {1},
	urldate = {2024-11-30},
	journal = {Physical Review A},
	author = {Gili, Kaitlin and Sveistrys, Mykolas and Ballance, Chris},
	month = jan,
	year = {2023},
	note = {Publisher: American Physical Society},
	pages = {012406}
}

@inproceedings{inajetovicEnablingNonlinearQuantum2023,
	address = {Berlin, Heidelberg},
	title = {Enabling {Non}-linear {Quantum} {Operations} {Through} {Variational} {Quantum} {Splines}},
	isbn = {978-3-031-36029-9},
	url = {https://doi.org/10.1007/978-3-031-36030-5_14},
	doi = {10.1007/978-3-031-36030-5_14},
	urldate = {2024-11-29},
	booktitle = {Computational {Science} – {ICCS} 2023: 23rd {International} {Conference}, {Prague}, {Czech} {Republic}, {July} 3–5, 2023, {Proceedings}, {Part} {V}},
	publisher = {Springer-Verlag},
	author = {Inajetovic, Matteo Antonio and Orazi, Filippo and Macaluso, Antonio and Lodi, Stefano and Sartori, Claudio},
	month = jul,
	year = {2023},
	pages = {177--192}
}

@article{ingelmannTwoQuantumAlgorithms2024,
	title = {Two quantum algorithms for solving the one-dimensional advection–diffusion equation},
	volume = {281},
	issn = {0045-7930},
	url = {https://www.sciencedirect.com/science/article/pii/S0045793024002019},
	doi = {10.1016/j.compfluid.2024.106369},
	urldate = {2024-11-30},
	journal = {Computers \& Fluids},
	author = {Ingelmann, Julia and Bharadwaj, Sachin S. and Pfeffer, Philipp and Sreenivasan, Katepalli R. and Schumacher, Jörg},
	month = aug,
	year = {2024},
	pages = {106369}
}

@inproceedings{kruseVariationalQuantumCircuit2024,
	address = {Rome, Italy},
	title = {Variational {Quantum} {Circuit} {Design} for {Quantum} {Reinforcement} {Learning} on {Continuous} {Environments}:},
	isbn = {978-989-758-680-4},
	shorttitle = {Variational {Quantum} {Circuit} {Design} for {Quantum} {Reinforcement} {Learning} on {Continuous} {Environments}},
	url = {https://www.scitepress.org/DigitalLibrary/Link.aspx?doi=10.5220/0012353100003636},
	doi = {10.5220/0012353100003636},
	language = {en},
	urldate = {2024-11-30},
	booktitle = {Proceedings of the 16th {International} {Conference} on {Agents} and {Artificial} {Intelligence}},
	publisher = {SCITEPRESS - Science and Technology Publications},
	author = {Kruse, Georg and Drăgan, Theodora-Augustina and Wille, Robert and Lorenz, Jeanette Miriam},
	year = {2024},
	pages = {393--400}
}

@misc{ortaliKineticDatadrivenApproach2024,
	title = {Kinetic data-driven approach to turbulence subgrid modeling},
	copyright = {Creative Commons Attribution 4.0 International},
	url = {https://arxiv.org/abs/2403.18466},
	doi = {10.48550/ARXIV.2403.18466},
	urldate = {2024-11-30},
	publisher = {arXiv},
	author = {Ortali, Giulio and Gabbana, Alessandro and Demo, Nicola and Rozza, Gianluigi and Toschi, Federico},
	year = {2024},
	note = {Version Number: 1},
}

@misc{liPotentialQuantumAdvantage2024,
	title = {Potential quantum advantage for simulation of fluid dynamics},
	url = {http://arxiv.org/abs/2303.16550},
	doi = {10.48550/arXiv.2303.16550},
	urldate = {2024-11-30},
	publisher = {arXiv},
	author = {Li, Xiangyu and Yin, Xiaolong and Wiebe, Nathan and Chun, Jaehun and Schenter, Gregory K. and Cheung, Margaret S. and Mülmenstädt, Johannes},
	month = mar,
	year = {2024},
	note = {arXiv:2303.16550}
}

@misc{kocherlaTwocircuitApproachReducing2024,
	title = {A two-circuit approach to reducing quantum resources for the quantum lattice {Boltzmann} method},
	url = {http://arxiv.org/abs/2401.12248},
	doi = {10.48550/arXiv.2401.12248},
	urldate = {2024-11-30},
	publisher = {arXiv},
	author = {Kocherla, Sriharsha and Adams, Austin and Song, Zhixin and Alexeev, Alexander and Bryngelson, Spencer H.},
	month = apr,
	year = {2024},
	note = {arXiv:2401.12248}
}

@misc{houLatticeBoltzmannSubgrid1994,
	title = {A {Lattice} {Boltzmann} {Subgrid} {Model} for {High} {Reynolds} {Number} {Flows}},
	url = {http://arxiv.org/abs/comp-gas/9401004},
	doi = {10.48550/arXiv.comp-gas/9401004},
	urldate = {2024-11-30},
	publisher = {arXiv},
	author = {Hou, S. and Sterling, J. and Chen, S. and Doolen, G. D.},
	month = jan,
	year = {1994},
	note = {arXiv:comp-gas/9401004}
}

@misc{dendukuriDefiningQuantumNeural2020,
	title = {Defining {Quantum} {Neural} {Networks} via {Quantum} {Time} {Evolution}},
	url = {http://arxiv.org/abs/1905.10912},
	doi = {10.48550/arXiv.1905.10912},
	urldate = {2024-11-30},
	publisher = {arXiv},
	author = {Dendukuri, Aditya and Keeling, Blake and Fereidouni, Arash and Burbridge, Joshua and Luu, Khoa and Churchill, Hugh},
	month = mar,
	year = {2020},
	note = {arXiv:1905.10912}
}

@misc{bakkerQuantumCarlemanLinearization2024,
	title = {Quantum {Carleman} {Linearization} of the {Lattice} {Boltzmann} {Equation} with {Boundary} {Conditions}},
	url = {http://arxiv.org/abs/2312.04781},
	doi = {10.48550/arXiv.2312.04781},
	urldate = {2024-11-30},
	publisher = {arXiv},
	author = {Bakker, Bastien and Watts, Thomas W.},
	month = mar,
	year = {2024},
	note = {arXiv:2312.04781}
}

@misc{budinskiEfficientParallelizationQuantum2023,
	title = {Efficient parallelization of quantum basis state shift},
	url = {http://arxiv.org/abs/2304.01704},
	doi = {10.48550/arXiv.2304.01704},
	urldate = {2024-11-30},
	publisher = {arXiv},
	author = {Budinski, Ljubomir and Niemimäki, Ossi and Zamora-Zamora, Roberto and Lahtinen, Valtteri},
	month = oct,
	year = {2023},
	note = {arXiv:2304.01704}
}

@article{schuldEvaluatingAnalyticGradients2019,
	title = {Evaluating analytic gradients on quantum hardware},
	volume = {99},
	url = {https://link.aps.org/doi/10.1103/PhysRevA.99.032331},
	doi = {10.1103/PhysRevA.99.032331},
	number = {3},
	urldate = {2024-11-30},
	journal = {Physical Review A},
	author = {Schuld, Maria and Bergholm, Ville and Gogolin, Christian and Izaac, Josh and Killoran, Nathan},
	month = mar,
	year = {2019},
	note = {Publisher: American Physical Society},
	pages = {032331}
}

@article{ljubomirQuantumAlgorithmNavier2022,
	title = {Quantum algorithm for the {Navier}–{Stokes} equations by using the streamfunction-vorticity formulation and the lattice {Boltzmann} method},
	url = {https://www.worldscientific.com/worldscinet/ijqi},
	doi = {10.1142/S0219749921500398},
	language = {en},
	urldate = {2024-11-30},
	journal = {International Journal of Quantum Information},
	author = {Budinski, Ljubomir},
	month = feb,
	year = {2022},
	note = {Publisher: World Scientific Publishing Company}
}

@article{yepezQuantumLatticeGasModel2002,
	title = {Quantum {Lattice}-{Gas} {Model} for the {Burgers} {Equation}},
	volume = {107},
	issn = {1572-9613},
	url = {https://doi.org/10.1023/A:1014514805610},
	doi = {10.1023/A:1014514805610},
	language = {en},
	number = {1},
	urldate = {2024-11-30},
	journal = {Journal of Statistical Physics},
	author = {Yepez, Jeffrey},
	month = apr,
	year = {2002},
	pages = {203--224}
}

@unpublished{itaniLatticeBoltzmannLinear2023a,
	title = {Lattice {Boltzmann} is {Linear} for {Single}-{Phase} {Incompressible} {Fluids}},
	url = {https://rgdoi.net/10.13140/RG.2.2.18131.25122},
	language = {en},
	urldate = {2024-11-30},
	author = {Itani, Wael},
	year = {2023},
	note = {Publisher: Unpublished},
}

@book{lacatusSurrogateQuantumCircuit2025,
    title = {Surrogate {Quantum} {Circuit} {Design} for the {Lattice} {Boltzmann} {Collision} {Operator}},
    author = {Lacatus, Monica and Möller, Matthias},
    month = jul,
    year = {2025},
}

@article{pandey_perspective_2020,
	title = {A perspective on machine learning in turbulent flows},
	volume = {21},
	issn = {1468-5248},
	url = {https://www.tandfonline.com/doi/full/10.1080/14685248.2020.1757685},
	doi = {10.1080/14685248.2020.1757685},
	pages = {567--584},
	number = {9},
	journaltitle = {Journal of Turbulence},
	shortjournal = {Journal of Turbulence},
	author = {Pandey, Sandeep and Schumacher, Jörg and Sreenivasan, Katepalli R.},
	urldate = {2026-01-07},
	date = {2020-10-02},
	langid = {english},
}
\clearpage
\appendix
\section{Tables Summarizing the Training of the Different Collision Ansatz}

\begin{table}[!htbp]
    \centering
    \begin{tabularx}{\textwidth}{XXXXXXX}
             \# of Layers&Learning Rate&  Batch Size&  Sample Size&  \# of Epochs&  Processed Count (rounded to nearest $10^6$)&  Loss\\
             16&0.1&  262,144&  1,048,576&  10&  10&  0.08318\\
 & & & & 20& 21&0.03704\\
 & & & & 30& 31&0.01331\\
 & 0.01& 262,144& 1,048,576& 40& 42&0.6407\\
 & & & & 50& 52&0.08484\\
 & 0.001& 262,144& 1,048,576& 60& 63&0.3085\\
 & & & & 70& 73&0.04147\\
 & & & & 80& 84&0.001032\\
 & & & & 100& 105&0.000565\\
 20& 0.1& 262,144& 1,048,576& 10& 10&0.0073727\\
 & 0.01& 262,144& 1,048,576& 10& 10&0.002292\\
 & & & & 15& 16&0.04326\\
 & 0.001& 262,144& 1,048,576& 15& 16&0.006037\\
 & & & & 20& 21&0.008651\\
 & & & & 30& 31&0.004102\\
 & 0.0001& 262,144& 1,048,576& 40& 42&0.008136\\
    \end{tabularx}
    \caption{Parameters and loss of the training of SEL-CRY-Inverse-SEL collision ansatz with amplitude encoding}
    \label{tab:amp-nonsym}
\end{table}

\begin{table}[!htbp]
    \centering
    \begin{tabularx}{\textwidth}{XXXXXXX}
             \# of Layers&Learning Rate&  Batch Size&  Sample Size&  \# of Epochs&  Processed Count (rounded to nearest $10^6$)&  Loss\\
             10&0.1&  131,072&  2,097,152&  10&  21&  0.0294\\
 & & & & 20& 42&0.08731\\
 & & & & 30& 63&0.07067\\
 & & & & 40& 84&0.01783\\
 & & & & 50& 105&0.02498\\
 & & & & 60& 126&0.04654\\
 & & & & 70& 147&0.04027\\
 & & & & 80& 168&0.05175\\
 & & & & 100& 210&0.09092\\
 & 0.01& 131,072& 2,097,152& 80& 168&0.03919\\
 & & & & 100& 210&0.00588\\
 & & & & 200& 419&0.00261\\
 & & & & 300& 630&0.1717\\
 & 0.001& 131,072& 2,097,152& 300& 630&0.0044\\
 & 0.0001& 131,072& 2,097,152& 500& 1,049&0.00211\\
 20& 0.1& 64,536& 1,048,576& 10& 10&0.1077\\
    \end{tabularx}
    \caption{Parameters and loss of the training of $2^{nd}$ Order SEL-CRY-Inverse-SEL collision ansatz with amplitude encoding}
    \label{tab:amp-nonsym-2}
\end{table}

\begin{table}[!htbp]
    \centering
    \begin{tabularx}{\textwidth}{XXXXXXX}
             \# of Layers&Learning Rate&  Batch Size&  Sample Size&  \# of Epochs&  Processed Count (rounded to nearest $10^6$)&  Loss\\
             64&0.1&  262,144&  1,048,576&  10&  10&  0.03933\\
 & & & & 20& 21&0.005144\\
 & & & & 30& 31&0.001867\\
 & & & & 40& 42&0.004426\\
 & & & & 50& 52&0.004056\\
 & & & & 60& 63&0.001527\\
 & 0.01& 262,144& 1,048,576& 70& 73&0.001962\\
 & & & & 80& 84&0.002453\\
 & & & & 90& 94&0.00472\\
 & & & & 100& 105&0.001967\\
 & & & & 200& 210&0.000396\\
    \end{tabularx}
    \caption{Parameters and loss of the training of SEL-CRY-Inverse-SEL collision ansatz with amplitude encoding and symmetry transforms}
    \label{tab:symamp}
\end{table}

\begin{table}[!htbp]
    \centering
    \begin{tabularx}{\textwidth}{XXXXXXX}
             \# of Layers&Learning Rate&  Batch Size&  Sample Size&  \# of Epochs&  Processed Count (rounded to nearest $10^6$)&  Loss\\
             8&0.1&  262,144&  1,048,576&  10&  10&  0.02103\\
 & & & & 20& 21&0.00189\\
 & & & & 30& 31&0.00185\\
 & & & & 40& 42&0.00213\\
 & & & & 50& 52&0.00101\\
 64& 0.1& 262,144& 1,048,576& 10& 10&0.03223\\
 & & & & 20& 21&0.00353\\
 & & & & 30& 31&0.00169\\
 & & & & 40& 42&0.00293\\
 & & & & 50& 52&0.00178\\
 & & & & 60& 63&0.00132\\
 & 0.01& 262,144& 1,048,576& 70& 73&0.00152\\
 & & & & 80& 84&0.00156\\
 & & & & 90& 94&0.00201\\
 & & & & 100& 105&0.00374\\
    \end{tabularx}
    \caption{Parameters and loss of the training of Strong collision ansatz with amplitude encoding and symmetry transforms}
    \label{tab:symstrongamp}
\end{table}

\end{document}